\documentclass[sigconf]{acmart} 

\setcopyright{acmlicensed} 
\copyrightyear{2026} 
\acmYear{2026} 
\acmDOI{XXXXXXX.XXXXXXX} 
\acmConference[CCS '26]{the 2026 ACM SIGSAC Conference on Computer and
Communications Security}{November 15--19, 2026}{The Hague, The Netherlands}  
\acmISBN{978-1-4503-XXXX-X/2018/06}  
\usepackage{booktabs}
\usepackage{longtable}
\usepackage{hyperref}
\usepackage{multirow}
\usepackage[inline]{enumitem}
\usepackage{xcolor}
\usepackage{tikz}
\usetikzlibrary{arrows}
\usepackage[draft,inline,nomargin,index]{fixme}
\fxsetup{theme=color,mode=multiuser}
\definecolor{op_bg}{RGB}{235, 245, 255}      
\definecolor{user_bg}{RGB}{250, 250, 250}    
\definecolor{border_gray}{RGB}{180, 180, 180} 

\newcommand{\scampost}[3]{%
    \par\noindent
    \ifx#1o%
        \def\tempbg{op_bg}%
    \else
        \def\tempbg{user_bg}%
    \fi
    \fcolorbox{border_gray}{\tempbg}{%
        \begin{minipage}{\dimexpr\linewidth-2\fboxsep-2\fboxrule\relax}
            \vspace{3pt}
            \textbf{\small #2} \par \smallskip
            {\footnotesize #3} 
            \vspace{3pt}
        \end{minipage}%
    }\par\vspace{5pt} 
}

\newcommand{\secref}[1]{\S~\ref{#1}}

\newif\ifcomment
\commentfalse

\ifcomment
\newcommand{\matt}[1]{{\bf \textcolor{teal}{Matt: #1}}}
\newcommand{\saleh}[1]{{\bf \textcolor{orange}{Saleh: #1}}}
\newcommand{\anish}[1]{{\bf \textcolor{brown}{Anish: #1}}}
\newcommand{\phani}[1]{{\bf \textcolor{red}{Phani: #1}}}
\else
\newcommand{\matt}[1]{}
\newcommand{\saleh}[1]{}
\newcommand{\anish}[1]{}
\newcommand{\phani}[1]{}
\fi

\usepackage{amsmath} 
\usepackage{algorithmic} 
\usepackage{array} 

\usepackage{dblfloatfix} 
\usepackage{url} 

\begin{document}
\title{Beyond the Prank: The Hidden Expertise of TSS Scambaiters}

  \author{Saleh Alsyefi}
  \authornote{Both authors contributed equally to the paper.}
  \affiliation{%
 	\institution{University of Bristol}
 	\city{Bristol}
 	\country{UK}}
  \email{saleh.alsyefi@bristol.ac.uk}

  \author{Anish Chand}
  \authornotemark[1] 
  \affiliation{%
 	\institution{Louisiana State University}
 	\city{Baton Rouge}
 	\country{USA}}
  \email{achan41@lsu.edu}

  \author{Matthew Edwards}
  \affiliation{%
 	\institution{University of Bristol}
 	\city{Bristol}
 	\country{UK}}
  \email{matthew.john.edwards@bristol.ac.uk}

  \author{Phani Vadrevu}
  \affiliation{%
 	\institution{Louisiana State University}
 	\city{Baton Rouge}
 	\country{USA}}
  \email{kvadrevu@lsu.edu}

\begin{abstract}

This research studies how Technical Support Scams (TSS) are being countered by a
uniquely dedicated community of volunteer counter-fraud operatives. Using a
careful subject selection strategy, we interviewed 17 individuals who actively
engage in TSS scambaiting activities in order to obtain insight into their
motivations, the operational methods of the scammers they combat, the
undocumented nuances of effective scambaiting action, and the various challenges
scambaiters face. In our analysis, we find a community rich not only with
insight into offenders, but with technical and operational expertise that is
often lacking in research efforts targeting these same populations.  At the same
time, we find key areas where the community could be better supported and
enabled. We discuss the implications of our findings for both future research
and community protection strategies.  

\end{abstract}

\ccsdesc[500]{Security and privacy~Social engineering attacks}
\ccsdesc[500]{Security and privacy~Social aspects of security and privacy}
\ccsdesc[300]{Human-centered computing~HCI design and evaluation methods}

\keywords{Technical Support Scams, Scambaiting, Social Engineering, Cybercrime}

\maketitle

%
\section{Introduction}
\label{sec:introduction}

Technical Support Scams (TSS) are a form of social engineering attack in which attackers impersonate technical support personnel and deceive victims into granting access to their computers or making fraudulent payments~\cite{liuUnderstandingMeasuringDetecting2023, miramirkhaniDialOneScam2017, vadrevu_what_2019}. Victims suffer especially from revictimization through refund scams, in which scammers use remote access to fabricate evidence of an over-refund and pressure victims into `paying back' funds~\cite{ic3_tech_support_scam}. TSS rely heavily on real-time audio conversations between victims and scammers, in interactions lasting around 17 minutes on average~\cite{miramirkhaniDialOneScam2017}. During these conversations, attackers build trust, make false technical claims, and guide victims through actions that ultimately enable financial theft.

Typically, users are exposed to these scams through a variety of well-documented online channels. Some well-known techniques include web search poisoning, malvertising, and email~\cite{srinivasanExposingSearchAdvertisement2018,liuUnderstandingMeasuringDetecting2023,YamadaIKKAK24}.
Collectively, these exposure channels form a flourishing ecosystem that enables the TSS threat actors to reach and monetize victims at scale. The severity of this is evident in the reported financial losses exceeding \$1.46 billion in the United States alone in 2024~\cite{fbi_ic3_2024}, marking a concerning continued upward trajectory. The sustained profitability of this ecosystem underscores both the effectiveness of the scammers' approach and the absence of effective defenses.

In the absence of comprehensive institutional and technical countermeasures, an online community of self-motivated individuals---commonly known as \emph{scambaiters}---has emerged. These internet vigilantes confront scammers directly, actively impersonating potential victims to engage, stall, and expose scammers. Scambaiters are often portrayed in popular media as merely wasting scammers' time or engaging in adversarial interactions purely for entertainment on social media platforms~\cite{rossHelloThisMartha2021,laato2020scambaiting}.

Consistent with this sentiment, this resource remains substantially underutilized by researchers. Liu et al.~\cite{liuUnderstandingMeasuringDetecting2023} built a TSS website detection system by passively collecting data from the Scammer.info forum ~\cite{noauthor_scammer_info_nodate}, demonstrating the potential value of leveraging community data. Similarly, several studies have analyzed publicly available scambaiting videos~\cite{woodAnalysisScamBaiting2023,berneyNavigatingShadowsCyber2024,rossHelloThisMartha2021} to gain limited insights into scammer-victim interactions. These approaches, however, are inherently constrained as these videos---often produced for entertainment---omit the crucial behind-the-scenes preparation and strategy development that speak to scambaiter goals and priorities, and reveal little about the post-call actions that form the meaningful output of many real scambaiting interactions.
In general, many of the resources used by scambaiters are overlooked by researchers studying the same offender communities that scambaiters target, leading to weaknesses in the operational deployment of novel countermeasures and data collection mechanisms. Similarly, the cybersecurity research community has given surprisingly little attention to how new technology can be leveraged to effectively support scambaiters as motivated and informed counter-fraud operatives.

In this paper, we present findings from a semi-structured interview study with 17 scambaiters, spanning a range of experience levels and engagement styles. Our participants include both everyday community members and a small number of highly-visible scambaiters with large public followings, including one individual with over a million followers. Our objective is to understand the scambaiting ecosystem from the perspective of its participants. Specifically, we examine scambaiters’ motivations, backgrounds, mental models, the planning and preparation they undertake prior to engagement, the technical and interpersonal skills they develop, and the ways in which they acquire knowledge. 
Through this perspective, we surface the practices and experiential knowledge of an understudied counter-fraud community and consider their implications for the different stakeholders involved in defending against and responding to TSS. We also identify the challenges participants encounter, along with their reflections on the community itself, including its perceived limitations and needs.

To the best of our knowledge, our work represents the first systematic effort to directly engage with scambaiters and harness their experiential knowledge. By doing so, we uncover rich, actionable insights into how defenders in practice learn about scams and counter scammers---insights that can directly help researchers in designing more effective technical defenses against technical support scams.
First, we find that the scambaiting community contributes to defenses against technical support scams in multi-dimensional ways. Beyond entertainment and simply wasting scammers’ time, many participants prioritize systematic information and evidence collection, with reporting as their primary objective. 
Second, we surface the persistent challenges faced by scambaiters in combating technical support scams, revealing gaps in existing technical and institutional responses to TSS.
Third, we translate our findings into actionable recommendations for a range of stakeholders, showing how scambaiter knowledge, resources, and practices can be leveraged to strengthen technical defenses, institutional responses, and broader scam-disruption efforts.

\section{Background and Related Work} \label{sec:related_work}
This research builds upon two primary streams of literature: studies on TSS and the emerging body of work on scambaiting activities.

\subsection{Understanding Technical Support Scams}
Technical Support Scams involve perpetrators who impersonate support agents from well-known technology companies (e.g., Microsoft, Apple) to deceive victims \cite{miramirkhaniDialOneScam2017}. Initial contact is often made through unsolicited channels such as fake pop-up browser warnings, cold calls, or deceptive emails, which create a sense of urgency and fear~\cite{liuUnderstandingMeasuringDetecting2023}. Scammers then employ a variety of social engineering tactics, authoritatively using confusing technical terms, to convince victims that their devices are infected with malware or suffering from errors, necessitating immediate and paid intervention.

Prior research has documented the operational aspects of TSS. Vadrevu et al.~\cite{vadrevu_what_2019} provided a large-scale analysis of social engineering campaigns, which include TSS, detailing the deceptive tactics used to lure and engage victims. Srinivasan et al.~\cite{srinivasanExposingSearchAdvertisement2018} focused on the infrastructure, particularly how scammers abuse search engine advertising and optimization to direct potential victims to their fraudulent services. Yamada et al. studied technical support scams targeting victims in Japan~\cite{YamadaIKKAK24}. The evolving nature of these scams, including the adoption of new technologies and payment methods, has also been a subject of recent studies \cite{acharya_conning_2024, acharya_scamchatbot_2024}, highlighting the technical sophistication and adaptability of TSS operators.

\subsection{The Phenomenon of Scambaiting}
Scambaiting is a fraud prevention activity in which individuals, known as scambaiters, intentionally interact with scammers without any intention of paying them. Scambaiters employ a broad range of tactics, including using virtual machines, fake personas, and custom scripts, to waste scammers' time, gather intelligence on their operations, and publicly expose their methods, often through online forums or video platforms, where interactions are routinely documented for education and entertainment of the public.

The academic literature presents a range of views on this activity. Economic analyses have suggested that scambaiting activity can be an effective countermeasure through the creation of false positives in a business model that relies heavily on self-selection of victims~\cite{herley2012nigerian}, but several commentators have raised concerns about the various motives of scambaiters and the legality of some actions considered under the umbrella of scambaiting~\cite{zingerle2013humiliating,dynelYouDontFool2021,419digilantes} or more broadly questioned the desirability or nature of vigilante action in online crime prevention~\cite{button2021exploring}. Although the broader impact of scambaiting is difficult to quantify, publicly reported cases describe scambaiters assisting law enforcement in high-profile investigations and takedowns of scam operations~\cite{sb_real_case1,sb_real_case2,sb_real_case3,sb_real_case4}.

 The outputs of scambaiter interactions have been adopted in past studies and analyzed for what they can reveal about offender strategies~\cite{woodAnalysisScamBaiting2023,rossHelloThisMartha2021}. As the entertainment motive of publicized scambaiting interactions can complicate such analyses, some researchers have attempted to automate scambaiting activity to provide experimental evaluations of scambaiting strategies or scammer behaviors~\cite{chen_active_2023,siadati2025send}. Such automation is also increasingly being adopted by scambaiters themselves~\cite{kitboga_ai}, raising questions about the evolving nature of scambaiting, which we hope in part to answer within the remainder of this paper.

\section{Methodology}\label{sec:methodology}
We used semi-structured interviews to qualitatively examine the lived experiences, technical tactics, and motivations of TSS scambaiters.
This section details our participant recruitment strategy, fraud mitigation measures, and our thematic analysis process.

\subsection{Participant Recruitment}

We recruited participants primarily from established online scambaiting communities identified through keyword searches and consultation with active community members. Our recruitment covered three main channels: \begin{enumerate*}
        \item \textbf{Specialized Forums \& Discord:} We posted recruitment calls on \textit{Scammer.info}\footnote{\textit{Scammer.info} has $\approx$24,000 members in their Discord server and forum} and \textit{TechScammersUnited}. These platforms are considered high-quality sources as they are dedicated specifically to scambaiting operations.
        \item \textbf{Reddit Communities:} We targeted high-traffic subreddits hosting scambaiting content, such as \texttt{r/ScamNumbers} (54k members) and \texttt{r/scambait} (586k members).
        \item \textbf{Direct Contact:} Establishing community trust was crucial for effective recruitment. Many scambaiters were either skeptical of researchers or wary of sharing their tactics publicly. To overcome this, we used gatekeepers: we contacted an administrator of one large community, who facilitated introductions. We then contacted all administrators/moderators of the targeted communities to seek permission for recruitment posts and gain their support. In one distinct case, we directly contacted a scambaiter celebrity (P19) with a mass following ($>$1M subscribers) to gain insights from a professionalized perspective.
\end{enumerate*}
To motivate participation, we offered \$30 USD in compensation for interviews\footnote{\$30 USD compensation via Tango Reward Link for US participants or region-specific Amazon gift cards for international participants.}. All procedures were approved by our Institutional Review Board (IRB).

\subsection{Screening and Fraud Mitigation}
\label{ssec:screening-and-fraud-mitigation}
\anish{remove trailing spaces}
Offering money for interviews can create a problem as it attracts bots, fraudulent respondents, and opportunists with no expertise in the topic. This is known as survey fraud~\cite{Zhang22, NguyenSAPV24}. We expected this to happen, so we built a screening process to filter the imposters and identify genuine experts. We filtered our original 146 applicants using three strategies discussed below.\anish{changed from list to save space}

\textbf{Blind Topic Recruitment.}
Hiding specific inclusion criteria in research allows for more participants to join without bias and prevents opportunistic participants from tailoring responses to the study's requirements. For this reason, we did not explicitly disclose that the study focused solely on TSS. Instead, we framed the study as being about scambaiting broadly, stating \textit{``We are conducting research on scambaiting and are looking for your help to participate in an interview''}. This hindered participants from guessing the selection criteria, ensuring that reported expertise was genuine.

\textbf{Consistency and Competence Filtering.}
Real experts have specific skills, while survey fraudsters commonly claim broad expertise to qualify for inclusion. To distinguish responses, we enforced a \emph{Channel Consistency Check} related to our undisclosed topic criteria. We automatically rejected participants who did not report scambaiting experience via both ``Voice phone call'' and ``Screen sharing/remote desktop connections'', as these are the consistent communication channels in TSS~\cite{miramirkhaniDialOneScam2017}. We also rejected 3 respondents who selected all available options, claiming expertise on every scam channel. This pattern is indicative of an attempt to manipulate the survey for compensation. Following our consistency filtering criteria, 39 candidates would be considered for interview.

\textbf{Qualitative Override.}
Strict binary rules can sometimes reject good participants by mistake. A candidate may click the wrong box, miss a question, or misinterpret a question. To overcome this, we manually examined written answers. If a participant failed the checkbox test but wrote text that showed specific knowledge of TSS practices or tools (e.g., referring to ``ConnectWise'', a screen sharing tool common in TSS), we overrode the filter to include them. This manual review ensured we retained knowledgeable applicants that would have been excluded in a strictly automated approach. In total, we identified 59 qualified participants (as shown in Tables~\ref{table:channel_stats} and \ref{table:screening_process}).

Following filtering, qualified participants were invited to interview. We observed a substantial participant dropout rate. Many (37) participants who applied and passed the screening stages did not reply to the invitation for an interview. Others (4) scheduled interviews with the research team, but did not attend. These behaviors are indicative of the deeply suspicious attitude of the community (see Figure \ref{fig:forum_post_thread} in the Appendix). Interviewees indicated that the candidates who did not show up likely feared that \textit{we} were scammers posing as researchers to unmask them. Others may have been deterred by the legal risks of scambaiting, fearing that talking to us might expose their vigilante activities to law enforcement scrutiny.

To test our approach, we deliberately interviewed two participants (P11, P12) who did not meet our criteria. As expected, these participants turned out to be survey fraudsters, providing vague, inconsistent, and often abnormally delayed responses during telephone interviews. Although excluded from the final dataset, both were compensated. We note that researchers can carefully design their recruitment using a similar selection strategy to overcome challenges of reliable participant validation, especially in domains vulnerable to abuse.
Ultimately, 18 participants from the screening pool and 1 external high-profile participant were interviewed, totaling 19. However, after filtering out the two deliberately included test participants (P11, P12) during the “Post-interview filtering” step shown in Table \ref{table:screening_process}, we were left with 17 valid interviews.

\subsection{Data Collection}
It was important to create a safe environment for participants to share their experiences without fear of exposure or legal repercussions. We followed an IRB-approved protocol, which we discuss more in the \emph{Ethical Considerations} section ~(\secref{sec:ethics}). We conducted semi-structured interviews lasting approximately 60 minutes. Before each interview, participants provided their signed consent for their participation and for audio recording purposes. The audio recordings were manually transcribed and stripped of any identifiable information. To preserve anonymity, interviews were conducted via VoIP/video conferencing with cameras disabled, and participants were encouraged to use pseudonyms.

The interview protocol was divided into seven thematic sections: (1) Introduction \& Motivation, (2) Planning \& Preparation, (3) Tools \& Technologies, (4) Engagement Tactics, (5) Challenges, (6) Analysis \& Reporting, and (7) Best Practices. This structure ensured consistent coverage of core topics, such as the ``VM setup'' and ``voice changing'' tactics, while permitting deep dives into unique participant experiences. Appendix~\ref{appendix:interview-guide} provides the interview guide.

\subsection{Data Analysis: Thematic Analysis}
We analyzed the transcripts using \emph{codebook thematic analysis} with a hybrid deductive-inductive approach~\cite{braun2021thematic}. Two coders independently coded four randomly selected transcripts and collaboratively resolved inconsistencies to establish an initial codebook informed by inductive coding and our interview structure. The coders then applied the shared codebook across the first six transcripts (P01--P06), resolving inconsistencies and merging codes into stabilized themes. Throughout this process, the coders met several times to analyze their codes, discuss similarities, and resolve differences.

Both coders independently coded four additional transcripts. We introduced a section-specific scoping rule (see Table \ref{table:coding_scope}), restricting coders to a subset of relevant categories for each interview section. This reduced the cognitive load of selecting from more than 150 codes and minimized out-of-context errors. We determined that for this exploratory qualitative study, conceptual agreement on identified segments was more critical than strict segmentation accuracy. Therefore, we calculated agreement only on units coded by both raters, treating silence as attentional variance rather than disagreement~\cite{STRIJBOS2007394}. This resulted in a Cohen's Kappa~\cite{mchugh2012interrater} inter-rater reliability (IRR) score of $\kappa \approx 0.84$ for the four transcripts.\footnote{Calculations available \url{https://github.com/anonymous-researcher-67/tss-interviews}} Although this approach inflates agreement scores, we used IRR strictly to assess conceptual agreement between coders over mutually identified coding segments given the large codebook. We clarify that the reported IRR statistic is not intended as a standalone measure of qualitative rigor or as exhaustive agreement over all possible codable spans. Instead, we consider the process of codebook discussion, refinement, and consensus as our primary indicators of rigor in this thematic analysis, where IRR is not always a necessity~\cite{mcdonald2019reliability}.

Following this consensus, the remaining transcripts were divided and coded between the two coders. We did not use saturation as a formal stopping criterion. Rather, our sample size was determined pragmatically by recruitment feasibility and participant availability, consistent with norms for qualitative interview studies~\cite{caine2016local}. Our only saturation-related observation was codebook stabilization. The final codebook can
be found in Appendices~\ref{sec:open_science} and~\ref{sec:codebook_appendix}.

\begin{figure}[t]
    \centering
    \includegraphics[width=\linewidth]{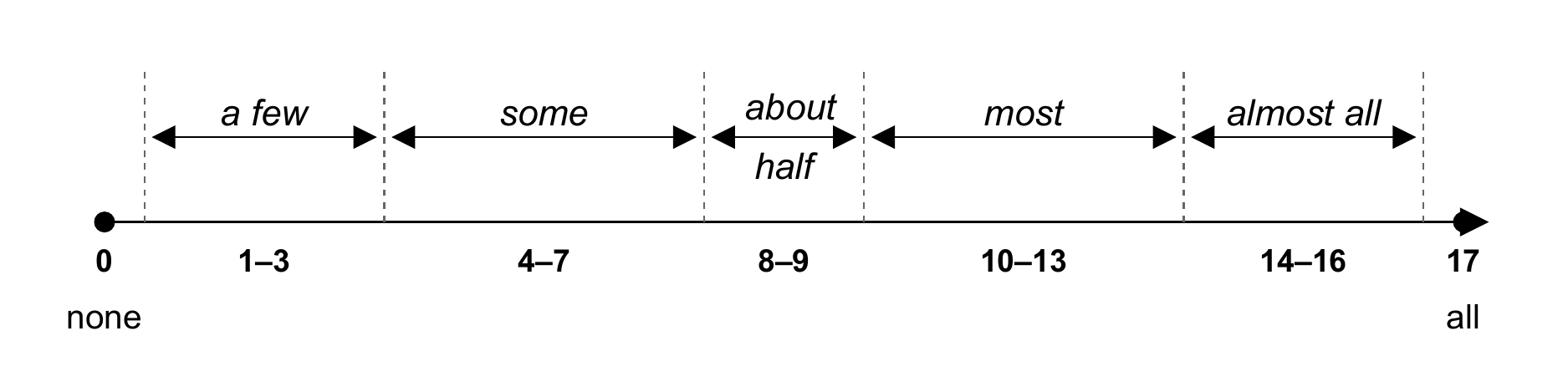}
    \Description{A scale showing the semi-quantitative descriptors used in the paper: a few, some, about half, most, almost all, and all, mapped to participant-count ranges from 1--3 up to 17.}
    \caption{Terminology used for participant-count ranges.}
    \label{fig:quantifier}
\end{figure}

Given the qualitative nature of our study, we avoid reporting exact participant counts. Similar to prior work~\cite{emami2019exploring,habib2020s}, where useful, we use consistent semi-quantitative descriptors as shown in Figure~\ref{fig:quantifier} to give a general idea about the prevalence of a finding.
These descriptors should not be interpreted as quantitative results.

\section{Scambaiting Motivations, Preparation and Risk Mitigation}
\label{sec:findings:s1}

In this section, we present the findings from our interviews covering the pre-engagement process of scambaiting. In particular, we present what drew participants to scambaiting, how they prepare, and how they manage the personal and technical risks involved.

\subsection{Motivations, Goals and Backgrounds}
\label{sub:motiv-goal-bg}
Participants described a diverse set of motivations that relate to both the operational goals of scambaiting interactions and the participants' varied educational and professional backgrounds.

\subsubsection{From Frustration to Engagement}
\label{sssec:from-frustration-to-engagement}
One entry point into scambaiting was strong anti-scam sentiment shaped by repeated exposure to scam attempts. Some participants described feelings of anger, frustration, or moral outrage triggered by persistent scams targeting themselves, their loved ones, or other vulnerable individuals, while some also framed their early engagement as a form of retaliation or emotional release in response to these experiences. As one participant explained, \emph{``It originally started as a form of revenge and a de-stressing way because I was getting frustrated with so many calls about extended warranties or Medicare coverage''} (P05). Beyond this, most described scambaiting as a means of resistance that also carried an element of enjoyment. For instance, P13 noted, \emph{``I see scam-baiting as a mix of civil disobedience\ldots but also I’m just having fun wasting their time.''} A common source of motivation was seeing online scambaiting content. Most participants reported that they saw \emph{``some videos on YouTube of Scammer Payback and Jim Browning and it got me interested in the whole thing''} (P10), consistently citing prominent creators such as Jim Browning, Kitboga, and Scammer Payback as the catalyst for their own involvement in scambaiting.
Despite these partially recreational motivations, participants emphasized that their engagement was not limited to entertainment. Most framed scambaiting as a form of public service and community protection. Participants frequently cited concern for elderly and vulnerable individuals as a major motivation. For instance, P04 stated, \emph{``My primary motivation is community protection, especially for the elderly.''} P17 explained, \emph{``\ldots if I’m wasting a scammer’s time, that costs them money and saves other people's time from calling up that particular scammer to, you know, not get scammed.''} This framing positions scambaiting not just as entertainment, but as a protective intervention. 
Importantly, as we see in the subsequent sections, this protective motivation extends beyond participants' reasons for engaging in scambaiting and shapes how they set their goals, prepare for engagements, and conduct them.

\subsubsection{Operational Goals}
\label{sssec:goals}
The protective motivations described above translated into clear operational goals that shaped how participants conducted their engagements. Understanding these goals is important because what scambaiters considered valuable can inform concrete target outcomes for future defensive interventions. These goals, as described by the participants, fell into two primary categories: harm prevention and the collection of actionable intelligence for reporting.
Most scambaiters explicitly described harm reduction through time-wasting and disruption as a goal. As P03 explained, \emph{``My main goal is to waste their time so they can't scam real victims.''} Others described keeping scammers occupied as a deliberate strategy to reduce call volume to potential victims (P16, P17).
Most participants also emphasized structured information gathering and reporting. Rather than merely prolonging calls, they sought actionable financial, contact, and technical infrastructure information for reporting.
For instance, P03 stated their objectives included, \emph{``gathering information like AnyDesk IDs [session identifiers from remote-access software], Bitcoin wallets, or bank accounts that can be used to try to shut them down,''} while P06 noted, \emph{``The ultimate definition of a successful call is if I'm able to get bank account information including account number, bank name, beneficiary name, and address''}. This has implications for researchers designing automated scam engagement systems. Specifically, scambaiters' emphasis on intelligence collection suggests such systems should go beyond successful interaction and support collection of actionable evidence.
  
Almost all participants described forwarding collected evidence to law enforcement, carriers, banks, or community forums to enable takedowns or victim notification. This reporting-oriented goal often coexisted with time-wasting, forming a dual objective of disruption and intelligence extraction and handoff.
For a few participants, success extended beyond the act of reporting itself to include observable real-world outcomes, namely shutdown of phone numbers, termination of accounts, and action taken by relevant external entities, including law enforcement. One participant stated that bait \emph{``might not be totally successful until I get confirmation that police have gone after the company responsible''} (P05). Reported outcomes also included scammers being arrested or infrastructure being dismantled. These sentiments reflect a broader aspiration to translate scambaiting efforts into tangible impact.

\subsubsection{Educational and Technical Background.}
\label{sssec:education-technical-back}
Given the technical nature of TSS scams and the operational goals described above, it might be expected that engaging in scambaiting requires significant technical expertise. However, participants in our study showed varied educational and professional backgrounds. While some participants held college degrees in computer science or had substantial experience in cybersecurity, about half described entering scambaiting without formal technical training. 
These details suggest that scambaiting is more accessible than its technical nature may imply, which can be explained partly by the rich ecosystem of community-developed tools and resources that lower the technical barrier to entry, as we discuss in~\secref{sub:preparation-and-prep}.

As one participant explained, \emph{``This is the first computer I’ve ever owned myself\ldots So it's no technical background at all''} (P15). Despite this, such participants remained actively involved in scambaiting activities. Participants attributed this accessibility partly to the availability of shared, community-developed tools that simplified the technical requirements for scambaiting. They also described community-driven learning, informal mentorship, and peer support as important ways of gaining relevant skills. For example, P03 described how \emph{``I've had great people in the community help me gain knowledge, learn new things, and access useful tools.''} These two forms of support are also concretely visible in the public resources maintained by the community. For example, a participant pointed to one such community tool: a pop-up generator that surfaces malicious, in-the-wild TSS pop-up pages. Such existing tools could be adapted by researchers studying TSS~\cite{liuUnderstandingMeasuringDetecting2023}, reducing the need to develop comparable functionality from scratch. We also observed a parallel form of learning support on Scammer.info, a community forum discussed by participants. For example, a popular megathread hosted on Scammer.info~\cite{noauthor_scammer_info_nodate} accumulates a rich collection of scambaiting resources and practical guidance accessible even without substantial prior technical experience. These resources provide an accessible entry point for security researchers to understand how scams and scambaiting operate in practice.

\subsection{Scambaiting Preparation and Setup}
\label{sub:preparation-and-prep}

Participants described a range of preparatory practices undertaken prior to engaging with scammers, covering their victim persona design as well as technical deception across virtualization, audio, and payment systems.

\subsubsection{Preparing Victim Personas}
\label{sssec:preparing-victim-personas}
To sustain engagements necessary to waste scammers' time and collect useful intelligence, participants needed to present themselves as genuine victims.
For this, about half of the participants commonly described actively preparing victim personas before engaging scammers. For example, one participant stated, \emph{``Before I even start, I come up with a name, a fake name, a fake persona, something that I'm going to be telling the scammer”} (P15). 
Prior work has similarly prepared hypothetical victim personas before contacting scammers, including biographical and contact information~\cite{YamadaIKKAK24}. Our participants described how this practice was operationalized across repeated scambaiting engagements by maintaining multiple personas and the supporting details needed to keep them credible.
P10 mentioned, \emph{``I have multiple personas and I've created a cheat sheet for each one that has their fake birthday, their fake address, their fake Social Security number, fake credit card numbers, and these are completely made-up numbers.''} 

A few participants also used voice modification in persona design. One participant also described modifying voice characteristics to align with their different personas, \emph{``I have voice changers to create different personas like an old woman, old man, or young ditzy girl”} (P04). Voice modulation has also been used in prior TSS research to conceal investigators' identity~\cite{YamadaIKKAK24}. 
Maintaining consistency across these details can become challenging, particularly when managing multiple personas, as one participant noted, \emph{``A significant ongoing challenge is keeping track of all my fake personas. When you create a fake world with addresses, voices, phone numbers, emails, and bank accounts, everything has to be consistent''} (P04).
Together, these strategies provide a picture of the information and supporting details that experienced scambaiters prepare before a scam engagement. This knowledge can inform security researchers conducting similar engagements or designing automated scam engagement systems for persona construction rather than rediscovering them through trial and error.

\subsubsection{Isolated Execution and Environment Stealth}
A consistent theme across almost all participants was the use of isolated execution environments to contain risk during live engagements with scammers. Technical support scammers frequently attempt to obtain persistent remote access through tools such as AnyDesk or ConnectWise---a tactic documented in prior work~\cite{miramirkhaniDialOneScam2017, YamadaIKKAK24}. In response, participants overwhelmingly relied on virtual machines (e.g., VMware, VirtualBox) and, in some cases, dedicated hardware systems to prevent compromise of personal devices, often combined with network-level protections such as VPNs. As one participant described, \emph{``I use a VPN to hide my real IP address, and I only let scammers connect to my virtual machine, never my real system''} (P03).

However, participants emphasized that isolation alone was insufficient because scammers may inspect the environment for signs of virtualization during remote sessions, to distinguish genuine victims from scambaiters. This dynamic is similar to sophisticated evasive malware that identifies signs of virtualization to suspend its malicious behavior to avoid dynamic analysis~\cite{kondracki_droid_2022}. Prior work has similarly shown that realistic virtual environments require deliberate modification, such as adjusting Registry keys, installing applications, and generating usage artifacts to resemble genuine user systems~\cite{miramirkhaniDialOneScam2017}. 
Consistent with this, most participants described many of these same practices, including modifying virtualization indicators and generating system usage activity such as browsing history, signing into disposable accounts, and adding photos or documents.
These artifacts were often tailored to match the chosen victim persona, such as an elderly user with recipes and family photos, tying environmental realism directly to their persona~(\secref{sssec:preparing-victim-personas}).

While prior work has touched upon some of these practices such as hiding virtualization artifacts and generating usage activity~\cite{miramirkhaniDialOneScam2017}, some of our participants described many additional mechanisms including simulating expected hardware and retaining control during remote engagements. P04, for example, configured \emph{``a fake webcam that shows a looped video of a notepad, since scammers get suspicious if a `laptop' has no camera.''} This shows how the absence of expected hardware can itself become an indicator of an artificial environment that must be accounted for. Similar mechanisms included simulated peripherals such as adding fake printers, and safeguards that allowed scambaiters to retain control when scammers attempted to block user input or obscure the screen.

Notably, some participants did not always perform these modifications themselves. Instead, they relied on pre-configured virtual machine images shared within the community that already incorporate these realism-enhancing modifications. As P14 noted: \emph{``There's a lot of preset virtual machines that you can basically download from other scambaiters which are great because then there's hidden that it is a virtual machine and stuff like that.''} Rather than requiring each scambaiter to independently discover and implement these changes, these pre-configured images package accumulated community knowledge into standardized bait setups. These setups lower technical barriers for less experienced scambaiters while housing \emph{protection} and \emph{realism} modifications that extend beyond the customizations seen in prior work~\cite{YamadaIKKAK24,miramirkhaniDialOneScam2017}. For security researchers conducting similar TSS engagements, these community-developed environments can provide a source of adversarially tested setups that can be adapted without requiring the same practices to be independently rediscovered.

\subsubsection{Deception Tools for Payment}
\label{sssec:deception-tools-for-sustained-engagement}
Prior research has reached the payment stages of TSS but terminated calls with scammers at this stage, rather than proceeding further~\cite{miramirkhaniDialOneScam2017}. In contrast, most participants described constructing or using deception tools to sustain engagement through and beyond this stage without being detected or transferring real money. Progressing beyond the payment stage is particularly important because it can yield actionable evidence for subsequent reporting and disruption. To successfully navigate the payment stage, participants mentioned using several clever, purpose-built tools, namely fake banking portals~\cite{bank_github}, gift-card generators, fabricated invoices, and fake cryptocurrency wallets. 
As one participant explained, \emph{``If they want bank account details, I have a fake bank account set up and use that''} (P13). These banking environments are more than simple static imitation pages. They incorporate interface and interactive behavior of online banking sites such as customizable bank themes, simulated accounts and transfers, transaction histories, and configurable login mechanisms such as two-factor authentication~\cite{bank_github}. A few participants also described maintaining broader suites of artifacts including realistic transaction histories, fake Amazon accounts with purchase records, and gift-card generators, to withstand deeper scrutiny during extended interactions (P03). This extended beyond digital artifacts to physical materials as well. One participant reported using physical props---such as fake gift cards, fake cash, or identification documents---to satisfy scammer verification demands during payment stages.

Interestingly, beyond the physical props, some participants also prepared for payment scenarios that moved outside the computer itself. Some scammers direct victims to physically travel to a bank and withdraw cash to complete a payment. P06 described preparing \emph{``cash withdrawal proof''} through simulating the physical environment of the driving trip by \emph{``creating sound effects like car horn noises when they ask for verification that I'm driving''}. P09 described a similar scenario requiring a store trip to purchase gift cards. To make this convincing, they took \emph{``them [scammers] on a `trip' to the store using sound effects for driving, walking, and store ambiance,''} going as far as to \emph{``even act as the store clerk.''} These practices show that sustaining engagement through the payment stage can require consistency not only across the victim persona and computing environment, but also across the financial and physical scenario presented to the scammer. Automated scam engagement systems can adapt these mechanisms to extend engagements beyond the payment stage where actionable evidence can be elicited. This is also valuable to law enforcement, as such evidence can support investigations and disruption efforts (see \secref{sec:discussion.le}).

\subsubsection{Hardware Setup.}
Prior TSS research limited themselves to virtualized and software-based environments when discussing scambaiting setups~\cite{miramirkhaniDialOneScam2017,YamadaIKKAK24}. Our interviews, however, revealed situations where software environments alone were insufficient to maintain realism, leading scambaiters to incorporate dedicated hardware into their setups. A key example of this was audio realism. Participants noted that real-time software voice changers often produced artifacts that reduced credibility during calls. As P04 explained, scammers \emph{``immediately get suspicious of certain voice changers, particularly ones that sound static-y''} noting that \emph{``so many scam baiters use the same `old lady voice'''}. Maintaining a voice that \emph{``does not sound like a robot”} was critical to maintaining credibility during extended calls (P16). As a result, some participants preferred dedicated hardware-based voice changers, namely GOXLR and Roland VT-4, which they found were more convincing. As P04 noted, \emph{``Physical voice changers like the Roland VT-4 or GOXLR are essential—there's no software that's as good yet for creating convincing voices.''}

\subsection{Protection and Risk Mitigation Practices}
\label{ssec:protection-and-risk-mitigation-practices}

Despite the safeguards provided by their engagement environments~(\secref{sub:preparation-and-prep}), almost all participants also treated separation between their real and scambaiting identities as an important protection against personal exposure. 
They used pseudonyms, anonymous email accounts, disposable telephony, VPNs, voice changers, and accounts not tied to their personal identity. For instance, P02 explained, \emph{``I use pseudonyms and I use no account that's connected to my real name,''} and emphasized the need to make sure \emph{``\ldots your real life and your fictional accounts never interact.''} Similarly, P18 noted the vocal component of this: \emph{``I use a voice changer. In fact, I'm using it right now to hide my identity.''}

These precautions are not unwarranted and respond to concrete risks. Several incidents witnessed in the scambaiting community illustrate the range of potential retaliation. For example, a prominent scambaiter was targeted through a phishing attack that resulted in the temporary loss of access to their online accounts~\cite{jim_browning_channel_restored,youtube_jim_browning_incident}. There are cases where leaked phone numbers of scambaiters have been spoofed by scammers, leading to misattribution and unwanted calls from victims or other scambaiters~\cite{scambaiter_spoofed_number}. More severe forms of retaliation have also been reported, including incidents where scammers falsely reported emergencies to law enforcement (``SWATing''), resulting in police responses to the scambaiter’s residence~\cite{scambaiter_swatting_case}. Some participants similarly expressed concern about these risks. P03 explained using \emph{``a VPN to protect my real location and prevent scammers from finding my approximate location or attempting to `swat' me,''} while P08 cautioned, \emph{``If you use your main phone number, there's a high chance of being swatted.''} Taken together, these cases and participant concerns highlight the adversarial environment in which scambaiters operate, where exposure of personal information can lead to harassment, misattribution, or escalation. This helps explain why participants treated strict identity separation and anonymity as necessary safeguards rather than routine privacy precautions. These same risks are relevant to security researchers who directly engage scammers, as such interactions may expose them to similar adversarial attention~\cite{harrasmentResearchers}.

\section{Findings: Active Engagement}\label{sec:findings_s2}
When scambaiters connect with a scammer, the active engagement phase begins. The main goal here is to play the role of a vulnerable victim while safely gathering information. To achieve this, baiters use specific methods to keep the scammer talking, avoid detection, and collect useful data for authorities.

\subsection{Target Sourcing}\label{subsection:target_sourcing}

Before an engagement can begin, scambaiters first need to identify an active scam operation and a means of contacting it. Finding active targets is a major hurdle; therefore, participants described using both community sources and original discovery mechanisms.

\subsubsection{Community and Peer Sourcing}\label{paragraph:community_and_peer_sourcing}
Public communities are a key scambaiting resource where scam phone numbers and websites discovered by community members are shared and verified. Existing work has demonstrated the research value of such forums by incorporating community data into TSS detection pipelines~\cite{liuUnderstandingMeasuringDetecting2023}. P03 reinforced such value, \emph{``For numbers from scammer.info or techscammersunited.com, they've already been verified as scammer numbers by the community''}. This exchange of leads also occurred through less public channels. Some participants relied on private peer networks to share active numbers. P09 reported, \emph{``A group of us use Slack to share numbers; it's a private community''}. Leads also entered these networks through personal contacts for some. P06 stated, \emph{``I also have family and friends report numbers to me''}. For security researchers, these community-sourced leads represent an underutilized source of real-world scam data that can bootstrap measurement, detection, and defensive research efforts.

\subsubsection{Technical Discovery and Passive Honeypots}\label{paragraph:technical_discovery_and_passive_honeypots}
About half of the participants used automated tools to scrape the web for open scam operations, using methods reported in previous studies~\cite{srinivasanExposingSearchAdvertisement2018}. P04 described finding tech support sites by searching \emph{``\nolinkurl{`site:sites.google.com'} and adding terms like `Microsoft' and `support' in quotations to find scammer sites''}. Using honeypots was similarly common, with about half of the participants describing setups similar to strategies used in existing research~\cite{prasad2023diving,vadrevu_what_2019}. P02 explained, \emph{``I have what's known as honeypot accounts\ldots there are guestbooks where you might post and scammers collect it''}. One participant also conducted scambaiting as part of their professional duties, using privileged resources available in that role. P07 noted, \emph{``Because I worked at the phone company, I could make SQL queries against phone records in order to use data analysis to find the scammer's phone number''}.

\subsection{Scammer Verification \& Evasion Tactics}\label{subsection:scammer_tactics}

Participants described scammers using several screening and evasion tactics to distinguish scambaiters from genuine victims. These tactics are relevant to the design of automated scam engagement systems that must account for the same screening checks that human scambaiters encounter. These are also relevant to law enforcement when investigators verify reported scam operations by directly calling suspected numbers. 

One screening signal was the caller's phone number provider. Some participants observed that numbers from popular services such as Google Voice or TextNow were sometimes ignored by scam call centers, motivating shifts toward alternative VoIP providers or PBX-backed numbers: \emph{``Scammers seem less likely to respond to Google Voice calls and services like TextNow—a lot of call centers filter those out''} (P06). Participants therefore had to ensure that the phone numbers they used appeared consistent with those of genuine victims, as scammers likely filtered incoming numbers even before engaging in calls.

\begin{figure}[t]
    \centering
    \includegraphics[width=\linewidth]{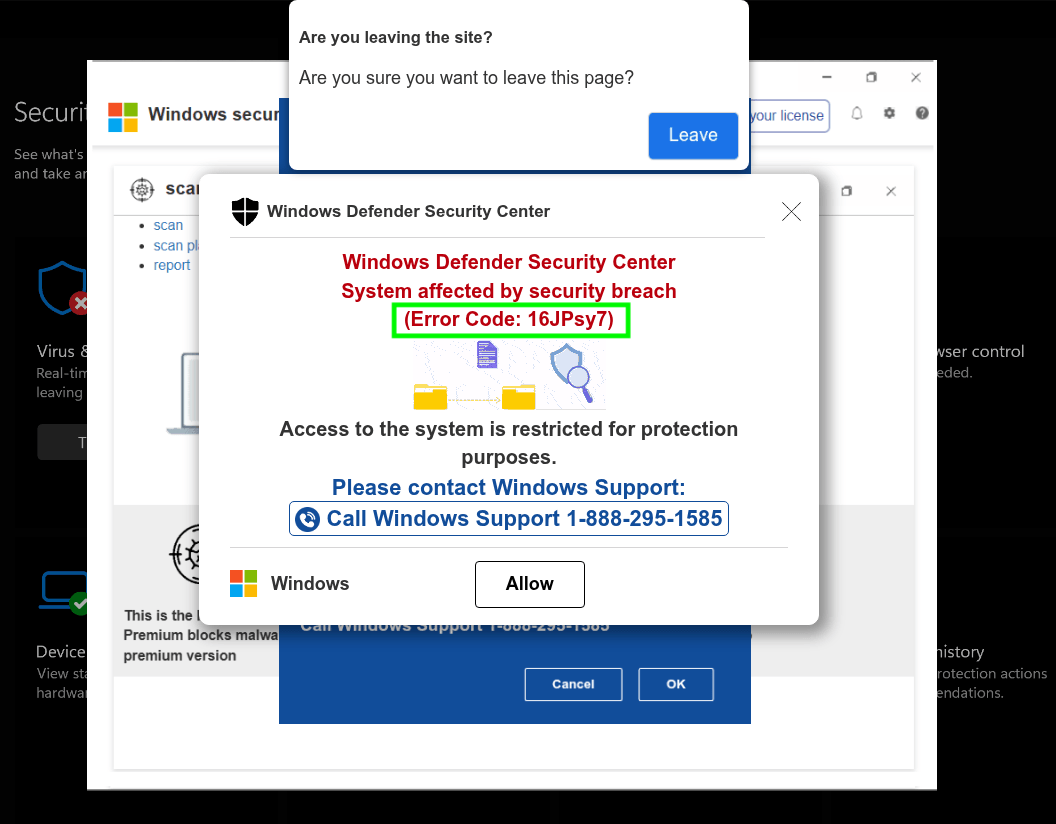}
    \Description{A technical support scam pop-up displaying a "Windows Defender Security Center" warning that the system is affected by security breach with the error code "16JPsy7". The pop-up states that the access to the system is restricted for protection purposes, and the victim needs to contact the fake Windows support using the toll-free number "1-88-295-1585".}
    \caption{A technical support pop-up containing the alphanumeric code that scammers can use to screen callers. The code is highlighted in green.}
    \label{fig:scammer-popup}
\end{figure}

Screening also occurred once a call began. Some participants described scammers using alphanumeric codes displayed on fake pop-up messages (Figure~\ref{fig:scammer-popup}) to screen incoming callers. Scammers can ask callers to provide the specific code to determine whether they had actually encountered the scam page. P08 noted, \emph{``they're looking for \ldots an ID they have in there to determine what's a scambaiter and what's not.''} While past research has captured these alphanumeric codes through screenshots of TSS warning pop-ups~\cite{YamadaIKKAK24,miramirkhaniDialOneScam2017,srinivasanExposingSearchAdvertisement2018}, to the best of our knowledge, ours is the first to identify their use as a caller-screening mechanism. As another screening tactic, scammers used regional information, checking whether a provided ZIP code was consistent with the phone number being used. As P08 warned, \emph{``the scammers will ask `Sir, can you please confirm your ZIP code'\ldots if you give them any that's not some way linked to the phone number you've got, they're very likely to hang the call up.''} Even when a caller fails this test and the call is dropped, scammers still find a way to benefit from it. The scammers harvest these active incoming numbers to sell to other criminals~\cite{liuUnderstandingMeasuringDetecting2023}. P10 found that \emph{``even if it's a bad number, they'll add it to their list because they want to sell a massive list to another call center''}.


Most participants also described rapid phone-number rotation as an evasion tactic. P06 explained, \emph{``Another factor is that there are many scam-baiters now, so call centers constantly change numbers and rotate through them. Numbers are often active for less than a day, making it difficult to find currently active numbers.''} This rapid turnover helps explain participants' reliance on community-sourcing as a steady stream of newly identified and verified targets~(\secref{paragraph:community_and_peer_sourcing}). Rapid rotation also matters beyond scambaiting. Security researchers designing defensive systems that rely on reported phone numbers must account for numbers being replaced quickly, while law enforcement and infrastructure and service providers may have limited time to verify reports and act before scammers move to new numbers.

\subsection{Crucial Conversational Techniques}\label{subsection:evasion_techniques_and_handling_payments}

In active conversation with offenders, scambaiters need to use social skills to manage forceful
social engineering techniques, overcome scammer demands for payment, and handle
call termination.

\subsubsection{Deflection and Interaction Techniques}\label{paragraph:deflection_and_interactions_techniques}

Participants described using a range of techniques to manage the `hard sell' tactics deployed by scammers. One strategy was to simply act confused to waste the scammer's time.
This mirrors the unexpected success of scripted phone bot, Lenny~\cite{sahin_using_nodate}, which has been shown to keep telemarketers engaged for an average of over 10 minutes by playing audio loops of a confused older man. As P13 explained, they \emph{``try to gaslight them into thinking I'm just some old grandma''}. To maintain the act under pressure, most participants relied on their personas. P04 noted, \emph{``If I'm playing an elderly person, I'll say things like `Who do you think you're talking to? You're talking to an old woman! Would you talk to your mother like that?'''}. Integrating these human deflection techniques can significantly improve the conversational resilience of automated scambaiting systems.
When the scammer tries to use fear, the baiter might just brush it off completely.
P15 stated, \emph{``I just laugh because I know they can't do anything and they're trying to take control''}.

\subsubsection{Payment Handling and Excuses}\label{paragraph:payment_handling_and_excuses}

A critical moment in any scam call is when the scammer demands money.
If a baiter fails to handle their demands convincingly, the scammer will immediately terminate the call and likely block their number.
While participants prepared the deception tools and props described in~\secref{sssec:deception-tools-for-sustained-engagement} for this stage, they also relied on conversational excuses to delay or redirect payment demands. To stall the process, some participants described pretending to lack technical knowledge. P04 explained, \emph{``I act completely naive and technically illiterate, asking lots of basic questions like `How do I download this program?'''}. If the scammer asked them to log into a bank, most participants invented excuses, as P05 explained, \emph{``Sir, that's not possible because my bank is a small local bank. They don't have websites or online accounts''}. 

\subsubsection{Session Termination}\label{subsection:session_termination}
The way participants decided to end a session often depended on what they had accomplished during the call. Understanding these
termination decisions is relevant to the design of automated scam engagement systems, which will need to determine when an
engagement should end.
Participants described several approaches to intentionally ending an interaction. About half of the participants described ending an interaction without revealing themselves after successfully getting the information they wanted. P04 noted, \emph{``if I've collected the information I need, I'll typically 'ghost' them rather than revealing I'm a scam-baiter''}. Another approach was to end the call with a dramatic reveal, letting the scammer know they have been tricked. P10 explained, \emph{``I reveal that I've been scam baiting them for an hour and a half, two hours, and you can just hear the deflation''}. 
Alternatively, calls naturally ended in a standstill because a few participants simply refused to pay. P05 noted, \emph{``I keep going until they give up and say, `have a good day, Sir'''}.

Not all session endings are deliberate choices by scambaiters. Calls could also terminate when a scambaiter made a mistake and broke character. 
Additionally, a lack of coordination over public numbers means a call might drop because the scammer's phone line is overburdened by bait calls.
P08 explained that because scammers become wise to callers who do not sound like typical victims, or simply due to \emph{``the fact of the numbers had so much use that they no longer answer''}, calls drop unexpectedly.

\subsection{Intelligence Gathering and Reporting}\label{subsection:intelligence_gathering_and_reporting}

As discussed in~\secref{sssec:goals}, intelligence collection was an important
operational goal for participants because the information gathered during
engagements could later support reporting and scam disruption. Participants
described both extraction mechanisms and the artifacts they targeted.

\subsubsection{Intelligence Extraction Methods}\label{paragraph:intelligence_extraction_methods}
Participants used several methods and tools to collect evidence and logs during engagements, ranging from basic note-taking to advanced technical capture. Some participants described, at a basic level, logging phone numbers, maintaining engagement notes, and behavioral notes to track scam patterns. At the more technical end, some scambaiters collected technical network data while engaging scammers. P06 explained, \emph{``I collect evidence through Microsoft Sysmon and TCP dump for full packet data''} similar to~\cite{YamadaIKKAK24}. A few participants also mentioned listening very closely to the background noise on the call to figure out exactly where the scam center is located. Such call-level cues may remain useful in scam detection even as scammers change the phone numbers and other infrastructure through which they operate. P10 described how they \emph{``dealt with a group of London-based scammers this past weekend that I was able to nail down to almost their neighborhood just by listening to them''}. 
To piece the whole puzzle together, some scambaiters also checked information gathered during engagements using external sources and reports from other scambaiters~\cite{dynelYouDontFool2021}. P05 stated, \emph{``I also look at reports from other people on the scam-baiting forum `scammer.info'''}. P08 added, \emph{``I'll look up who owns the phone number''}. These practices show how call-level and infrastructure signals can be useful in the design of scam detection systems. Such intelligence can also aid law enforcement in characterizing and corroborating scam operations (see \secref{sec:discussion.le}).

\subsubsection{Target Artifacts for Reporting}\label{paragraph:target_artifacts_for_reporting}
Through the extraction methods discussed above, participants collected several types of actionable information for reporting~\cite{buttonPolicingCrossborderFraud2025}. The financial information comprised several artifacts, namely money mule information (individuals or accounts used to launder stolen funds)~\cite{liuUnderstandingMeasuringDetecting2023}, scammer bank accounts, cryptocurrency wallets, MoneyGram, PayPal, and Zelle accounts.
P04 described collecting \emph{``Zelle accounts, money mule information, physical addresses''}, while P14 collected \emph{``PayPal, MoneyGram and Bitcoin wallets''}.
P02 reported, \emph{``I just collected, I think 73 bank accounts from a scammer''}. A few participants also inspected fake payment websites to identify ownership, with P05 noting that an online payment portal \emph{``reveals the company's name on the receipt''}. 

Beyond financial data, most participants also collected direct contact details for scam centers. P15 focused on \emph{``gathering phone numbers like callback numbers''}. On a technical level, about half of the participants captured the IP addresses and session identifiers for remote-control software for reporting. One participant even routed this intelligence to security experts, \emph{``I gather information about remote access tools scammers use and send that information about domains and access codes to gray hat security researchers''} (P06). These findings show that the scambaiting community can obtain several forms of actionable intelligence that can help law enforcement track and disrupt scam operations. If automated scam engagement systems can be designed to collect these same target artifacts, they could make such intelligence available to law enforcement at a much larger scale, potentially supporting more systematic investigation and disruption of TSS operations.

\section{Findings: Outcomes, Challenges, and Evolving Practices in Scambaiting}
\label{sec:findings:s3}
In this section, we present findings from our interviews regarding how scambaiters document their engagements, report findings to external entities, and evaluate the success of their activities.

\subsection{Engagement Outcomes and Practices}
\label{sub:engagement-outcomes-and-practices}
Participants described scambaiting as a structured process that can produce actionable artifacts that can inform reporting efforts and, ultimately, evaluate the success of an engagement.

\subsubsection{Documentation and Artifact Collection}
\label{para:documentation-and-artifacts-collection}

Any live TSS interactions are comparatively difficult for researchers to collect because they require sustained engagement with scammers. As such, it becomes difficult to develop and evaluate defensive systems in the absence of adequate datasets.
Our participants described routinely recording, documenting, and revisiting their own engagements. These artifacts provide a potentially valuable source of scammer interaction data that can inform defensive efforts.

Participants described using recordings and notes to reflect on past interactions, refine conversational strategies, and learn scam scripts. As P04 noted, \emph{``I mainly re-listen to recordings to refresh my memory on specific details or scripts''} while P15 explained, \emph{``So while I'm editing [video]\ldots I'm trying to figure out: OK, what could I have said there? What could I have done better here? How could I have stopped this from happening?''} Such records can provide security researchers with data to study scam scripts, strategies involved, and how scams change over time. However, researchers need to keep ethical considerations in mind, such as varying recording consent across jurisdictions.
Similarly, policymakers can use the scam playbook to design materials for scam awareness and public education.

Participants also mentioned using these artifacts in a collaborative setting, sharing recordings and observations with other scambaiters to receive feedback or stay updated on evolving scam techniques. P03 noted that most of their improvement derived from peer feedback rather than solo analysis, and described engaging with the community, \emph{``Sometimes I analyze my calls, but more often I'm in a Discord server voice channel with other scambaiters while making calls, and they give me guidance and feedback in real-time.''} This is consistent with our earlier observations highlighting community support for skill growth~(\secref{sssec:education-technical-back}).

\subsubsection{Reporting to External Entities}
\label{sssec:report-to-external-entities}

The protective motivations of scambaiters~(\secref{sssec:from-frustration-to-engagement}) led participants to report the collected intelligence~(\secref{paragraph:intelligence_extraction_methods}). However, TSS reporting is challenging as scams span financial, technical, and legal domains, making it unclear where different types of information should be reported. Participants had learned through experience how to route this information to banks, law enforcement, service providers, and community organizations. These reporting pathways are also useful for security researchers who may collect similar TSS intelligence and need to identify entities positioned to act on it.

Banks were a prominent destination for reporting financial intelligence. Bank reporting was often prioritized due to its potential for immediate impact. As one participant noted, \emph{``If I get bank account information, I contact those banks directly, usually by email''} (P06), while P02 emphasized that obtaining such details allows them to \emph{``get it reported to a bank''}. Participants highlighted that banks can act quickly, for example by freezing accounts or stopping transactions. Notably, participants also described establishing direct relationships inside financial institutions rather than relying exclusively on general reporting channels. P02 explained, \emph{``Typically in the anti-money laundering department of banks, [they] know me. I send them through these contacts.''} We hypothesize that these relationships suggest financial institutions may be receptive to intelligence generated by scambaiters as it affects their own customers and systems. Such relationships may also indicate collaboration opportunities for security researchers.

Participants also routed intelligence to law enforcement. Public reporting pathways included the Federal Bureau of Investigation's {Internet Crime Complaint Center (IC3)}, the Federal Trade Commission (FTC), and Homeland Security Investigations. P06 explained, \emph{``I forward the gathered information to the FBI's Internet Crime Complaint Center (IC3)~\footnote{https://www.ic3.gov} and the FTC~\footnote{https://reportfraud.ftc.gov/}.''} Reporting to law enforcement was framed as a longer-term or higher-level goal seen as a pathway for investigation or prosecution. Beyond these public channels, one participant had also developed direct law-enforcement contacts. P06 described an ongoing effort to establish connection, \emph{``I have a direct FBI law enforcement contact that I've met with in person, and we're currently working on potentially setting up a direct information sharing channel between myself and one of the FBI cyber centers, though that's still in progress.''} 

Reporting also extended to infrastructure and service providers responsible for resources abused during scam operations, namely telecom carriers, domain registrars, and online service providers. Reporting scam phone numbers to carriers was particularly common, with one participant explaining that they \emph{``report the scammers' phone numbers to their respective carriers''} (P05). They also described reporting domains, emails, and IP addresses to registrars or service providers, with the expectation that abuse teams would take action on malicious accounts or infrastructure.

Finally, reporting also extended into community and organizational channels. Participants shared intelligence within scambaiting forums, Discord servers, or with organizations involved in anti-fraud efforts. For example, P16 described collecting information and \emph{``taking that directly back to the forum and leaving it there''}, while others passed on information to specific organizations or other scambaiters with established enforcement contacts.

\subsection{Challenges and Constraints}
\label{sub:challenges}
Our data surfaces various technical and emotional challenges scambaiters face in their operations. We categorize these challenges into workload issues, responsiveness to reports, public awareness issues, and ethical and legal quandaries.

\subsubsection{Time Workload and Burnout}
\label{sssec:time-workload-and-burnout}
Given that participants mentioned wasting scammers' time as a primary goal and a key metric for success~(\secref{sssec:goals}), it is only natural that effective scambaiting sessions demand long sustained efforts. Prior work engaging with TSS scammers reported scammers taking an average of 17 minutes to reach the payment stage~\cite{miramirkhaniDialOneScam2017}, while~\cite{YamadaIKKAK24} reported an average call time of 87.3 minutes in a more recent work. As such, it becomes a huge challenge for these volunteers who have to balance scambaiting with their personal and professional obligations. The very mechanism through which scambaiting creates one of its valuable outputs---occupying scammers' time---also consumes scambaiters' time equally. P03 noted, \emph{``A typical successful call lasts around 2-3 hours.''} Another participant noted that they do not \emph{``always have 4 hours to sit on the phone as I'm a very busy person''} (P06). The imbalance itself is a challenge as P02 noted, \emph{``there are so many scammers and there's only so many hours in the day''}. Moreover, beyond time constraints, prolonged engagement with adversarial actors can take a psychological toll and lead to emotional burnout~\cite{scambaiter_mental_health}, as seen in other volunteer communities~\cite{dosono2019moderation, steiger2021psychological}.

\subsubsection{Delays and Routing Gaps in Reporting}
\label{sssec:delays_and_routing_gaps}
As discussed in~\secref{subsection:intelligence_gathering_and_reporting}, scambaiters gather scam intelligence to report to external entities for further action~(\secref{sssec:report-to-external-entities}). Collecting the scam intelligence is only half the challenge for scambaiters. Whether scambaiters could identify an appropriate recipient and whether the concerned entities can act upon it in time determines whether the intelligence leads to scam disruption. Participants described both of these as persistent challenges. Participants found it challenging to direct the gathered evidence to the correct entity. P15 described reaching a point where they thought, \emph{``Well, what's the point? I have nowhere to put, report this, so what's the point of gathering at all''}, before eventually receiving guidance from fellow scambaiters. 
Similarly, P14 described their struggle when calling a bank directly to report collected bank account details, only for the representative to respond, \emph{``I don't know what to do with this.''} 
P14 summed up his frustration: \emph{``I don't care who to report to\ldots I think that's really, really tricky to know what to do with this information and where to put it\ldots besides just calling them.''}

Even when the appropriate recipient was known, participants voiced concerns around the time required for institutions to act. This was especially the case for the law enforcement authorities, whom they considered too slow to act. One participant said, \emph{``For law enforcement to force banks to act requires warrants and subpoenas, which takes weeks or months''} (P06). 
Scammers, by contrast, rotate phone numbers and web domains rapidly to burn clear connections to criminal activity~(\secref{subsection:scammer_tactics}), meaning that some indicators may become stale before formal action is taken.

\subsubsection{Irresponsible Actors}
\label{sub:irresponsible}

Participants reported frustration with a variety of actors, with a focus on systemic accountability. A key concern was the role of telecom carriers in enabling scam operations, particularly due to weak verification and oversight mechanisms that allow scammers to easily obtain communication infrastructure. As one participant illustrated, \emph{``you cannot have a scammer clearly from India saying to a carrier, `Hey, I live in California. Can I please buy this big batch of phone numbers from you?', and the carrier responds with, `Yeah, sure'\ldots''} (P01). This sentiment of inadequate response has been experienced by others in the community too. For instance, in one well-documented case~\cite{carrier_inaction}, a baiter reported over 260 fraudulent numbers to a single telecom provider but described the response as ineffective: the provider representative labeled the activity as ``spoofing'' and instead directed the baiter to external authorities (FCC), while clarifying that only customers could formally report abuse on their network. 
Such inadequate responses are insufficient to meaningfully disrupt scam operations, and such experiences can lead to a decrease in motivation to report. As P19 noted, \emph{``I mean, I used to be really zealous about reporting and then nothing would happen. And so it, it took a lot of the wind out of my sails.''} Such lack of proactive enforcement not only fails to disrupt scam operations but can also lead to loss of willingness to report among scambaiters. \matt{Feels like this needs slightly more.}\anish{added a case and quote to expand this.}

\clearpage
\subsubsection{Public Awareness}
\label{sub:awareness}
Participants also stressed the importance of improving public awareness and victim support, often describing these as persistent challenges in combating scams. They emphasized that a lack of awareness and guidance creates vulnerability, as P02 noted, \emph{``What would help is more education. The big battle is getting other people to be aware of these scams.''} Similarly, P05 stated that addressing scams \emph{``mainly boils down to informing the public on what's going on with these scams and teaching them how to fight back.''} In response to these gaps in awareness, participants described actively trying to educate others about scam tactics, both by demonstrating how scams unfold and by explaining how to respond. For example, one participant explained using scambaiting interactions to \emph{``show everyone that what strategies they [scammers] use to get these people to get their money''} (P15) while P19 noted that their content creation serves an educational purpose. They also voiced the need for accessible resources about how to report incidents or seek help via official reporting channels. These perspectives reflect that participants view scambaiting not only as a way to disrupt scammers, but also as a means to educate potential victims and reduce overall susceptibility to scams.

\subsubsection{Ethical Risks and Legal Gray Areas}

Participants showed a strong awareness of the legal and ethical risks associated with scambaiting, often describing deliberate efforts to remain within legal boundaries while engaging scammers. They explicitly distinguished between acceptable and unacceptable behaviors, particularly avoiding unauthorized system access. As P18 summed up the distinction clearly, \emph{``\ldots simply calling and wasting time and gathering info on the phone is completely legal; hacking is not.''} P17 similarly recommended not doing \emph{``anything malicious like hijacking into a computer unauthorized.''} However, this distinction was not always clear-cut in practice. Participants acknowledged that scambaiting can involve legally ambiguous situations and perceived gray areas, especially given the cross-national nature of scams. For instance, P04 acknowledged some more aggressive practices:  \emph{``I don't believe I'm technically breaking any laws\ldots I know people who backdoor systems\ldots but I don't do that personally because it could potentially ruin my actual career.''}

The participants were also aware of ethical issues. For instance, P14 made sure \emph{``to double check if the address I use is not related to [a] real person,''} during their persona creation. Participants also reported avoiding doxxing or harassment (P13), and focusing on lawful information gathering and reporting to authorities (P05, P15). A few expressed stronger moral justification for their actions. For instance, P10 dismissed ethical concerns altogether, stating, \emph{``Ethically, I see zero issues. These people are thieves, simple.''} Participants also described community-driven norms that aim to reduce harm and enforce responsible behavior. These practices were particularly focused on protecting innocent third parties. For example, P15 explained that forum rules are designed to \emph{``weed out legitimate numbers and only focus on confirmed scammer numbers\ldots to prevent our users from unknowingly baiting innocent third parties.''} Overall, these accounts suggest that scambaiting operates within a spectrum of legal compliance, perceived gray areas, and community-enforced practices, with participants well aware of ethical and legal risks.

\subsection{Evolving Scam Ecosystem}
\label{sub:evolving-scam-ecosystem}

Participants described the scamming landscape as dynamic and evolving, with changes observed in scale, tactics, tools, and adversarial awareness. 
At the technical level, participants reported shifts in the tools and infrastructure used by scammers. Common remote access tools such as AnyDesk and TeamViewer are being replaced with more persistent and harder-to-remove alternatives. P04 explained that scammers have \emph{``switched to using ConnectWise ScreenConnect, which is much more insidious''}, while another noted the use of modified or specialized versions designed specifically for scamming, \emph{``Recently, many scammers are using a Russian fork of ConnectWise or ScreenConnect designed solely for scamming''} (P05). This enables continued access even after system restarts (P05), reflecting a shift toward more resilient infrastructure. In response to the potential misuse by threat actors, ScreenConnect removed customization features that had allowed bad actors to run the program without user knowledge, hide UI indicators, and impersonate legitimate businesses through custom logos and backgrounds~\cite{screenconnect_patches}. This response provides a concrete example of how infrastructure and service providers can reduce the misuse of their services through platform-level changes.

Participants reported an evolution in scam scripts and interaction strategies. Rather than relying on static narratives, scammers dynamically adapt their scripts during calls and blend multiple scam types. For example, one participant noted that a scam may begin as an Amazon purchase issue about an iPhone, but then shift to broader claims such as \emph{``we found multiple bank accounts registered under your name''} or fraudulent activity in other locations (P05). Similarly, another participant observed that \emph{``modern Microsoft scams often turn into banking pretexting scams''} where scammers claim foreign hackers have compromised the victim's accounts (P03).

Participants reported a major shift in payment methods, away from traditional mechanisms such as gift cards and towards cryptocurrencies. This matches prior work on TSS operations, which found that scammers' preferred payment modes included cryptocurrency alongside gift cards and PayPal~\cite{acharya_conning_2024}. One participant described this shift as \emph{``everything was gift cards''} in the past, whereas now victims are directed to \emph{``go to a Bitcoin ATM, put it in my wallet''} (P15). Similarly, another participant noted that scammers are \emph{``moving away from using mules and shifting more toward cryptocurrency''} (P06). This shift by scammers reflects an increased emphasis on anonymity and reduced risk of financial recovery.

Importantly, participants highlighted an emerging feedback loop between scambaiters and scammers. Scammers actively consume scambaiting content and adapt their behavior accordingly. As P02 noted, \emph{``a scammer\ldots can go see all the videos on YouTube''}, while another observed that scammers are becoming \emph{``way more attuned to the basic scambaiter''} (P16). This has led to increased awareness and defensive behavior, making it more difficult to deceive or engage them.

\subsection{Towards Automation and AI}
\label{sub:automation-ai}
Participants described automation in scambaiting ranging from simple scripted tools to emerging AI-assisted systems. At the basic level, some participants reported familiarity with \emph{Lenny}~\cite{lenny_origin, sahin_using_nodate}, a scripted phone bot that relies on pre-recorded audio responses. Its predictability was viewed as a limitation, with P08 noting that scammers can \emph{``catch on very quickly.''} Some participants also used large language models as supporting tools, particularly for persona development. For example, P13 used an LLM to generate contextual artifacts such as \emph{``a fake 80th birthday speech''} to make a persona more believable. These uses remained largely human-in-the-loop rather than fully automated.

About half of our participants were also aware of experimental efforts toward fully automated conversational scambaiting. Prior work has demonstrated the feasibility of LLM-based scambaiting against email scammers~\cite{bajaj_automatic_2023}. While most had not directly used such systems, they were aware of ongoing experimental developments within the community and among prominent scambaiters~\cite{kitboga_ai}. They also mentioned \emph{Daisy}, an AI ``granny'' bot developed by O2, a British telecommunications company, to engage scammers~\cite{o2_daisy}, using scam-interaction recordings from prominent scambaiter Jim Browning to produce realistic victim-like responses~\cite{jim_browning_daisy}. This further highlights the potential for integrating the scambaiting community’s accumulated data and experiential knowledge into the development of automated scam engagement systems.

Given these early signs of AI-driven conversations, most participants were receptive to incorporating AI into their scambaiting, although some were hesitant about replacing direct interaction. For instance, P03 preferred \emph{``the direct interaction and control that comes with handling calls personally, rather than delegating to AI tools.''} In contrast, a few participants envisioned more expansive uses of automation for disruption efforts. P19 described a \emph{``dream scenario''} in which AI systems could overwhelm scam operations: \emph{``I want to have 30 bots at all times talking to\ldots a call center that has 30 people,''} further expressing that governments, researchers, and scambaiters working together could have an \emph{``army of bots''} to overwhelm scam centers.

Participants' experiences also provide practical requirements for security researchers designing automated scam engagement systems. They emphasized grounding systems in real scammer scripts and interaction patterns. Beyond script awareness, most participants stressed conversational design as critical to sustaining engagement. Some participants stressed using diverse voice samples and carefully designing personas to align with specific scam contexts, maintaining deception throughout the interaction.
Consistent with the operational goals described in~\secref{sssec:goals}, most participants identified time-wasting as one potential indicator of success, but also emphasized actionable outcomes. Automated scam engagement systems should aim to collect and structure useful intelligence, such as payment details, phone numbers, or infrastructure indicators, which can then be reported or acted upon.

\section{Discussion and Recommendations} \label{sec:discussion}
Our findings show that most participants described prosocial motivations for scambaiting. Their workflow involved substantial preparation, varied engagement tactics, and considerable post-engagement effort. Participants developed or adapted tools---such as deceptive virtual environments and fake payment systems---and drew on shared resources to collect and report information for scam disruption. These activities also imposed considerable demands, with repetitive manual tasks contributing to burnout and reporting difficulties causing frustration.

We draw on these findings to outline recommendations for five primary stakeholder groups that occupy distinct but connected roles in responding to TSS.
With our recommendations, we hope to support safer scambaiting and help stakeholders leverage scambaiting community expertise for better defensive efforts.

\subsection{For Security Researchers}
\label{sec:discussion.sr}

\subsubsection{Detection and Mitigation Systems}

Given participants' detailed knowledge of current scam tactics~(\secref{subsection:scammer_tactics}), researchers designing detection and mitigation systems should utilize active scambaiting forums as a potential source of current intelligence on both established and evolving scam tactics and infrastructure. These resources can support measurement, detection, and defensive research~\cite{liuUnderstandingMeasuringDetecting2023}, while providing artifacts from stages of the scam lifecycle that researchers may not typically reach, including financial, technical, and infrastructure indicators collected during direct engagements. These artifacts can help researchers understand the full scam lifecycle and design defensive systems. As a precautionary measure, researchers should validate public data before incorporating it into their systems. They should also account for the dynamic nature of the scam ecosystem~(\secref{sub:evolving-scam-ecosystem}) by designing systems that are adaptive, resilient, and responsive to adversarial challenges rather than treating scam tactics or infrastructure as static.

\subsubsection{Automated Scam Engagement Systems}

Automated TSS scambaiting highlights both an opportunity and a challenge for security researchers. Importantly, these systems need to account for the screening and verification tactics that scammers use to identify scambaiters~(\secref{subsection:scammer_tactics}). Researchers could adapt the mechanisms and conversational techniques that scambaiters use to pass these screening checks, including caller-number verification, pop-up code screening, and regional consistency checks.

Sustainable engagement also requires a consistent and believable victim persona across the virtual environment, voice profiles, and contextual cues such as sound effects~(\secref{sub:preparation-and-prep}). 
The community's use of pre-built baiting images further highlights an opportunity for adapting standardized, defender-focused virtual environments. Similarly, while prior work has justifiably disengaged at the payment boundary~\cite{miramirkhaniDialOneScam2017}, scambaiters routinely cross it using purpose-built deception tools such as fake online banking sites~(\secref{sssec:deception-tools-for-sustained-engagement}). Researchers could incorporate such tools to continue engagement past payment stages without having to disengage or pay bad actors.

Finally, these systems should go beyond sustaining engagement and support the broader operational goals identified by participants, particularly the collection of actionable intelligence for reporting and scam disruption~(\secref{sssec:goals}).

\subsubsection{Design for Participant and Researcher Safety}
Participants treated identity separation as a necessary safeguard against retaliation from their adversary~(\secref{ssec:protection-and-risk-mitigation-practices}). Researchers operating in the same space should consider the possibility of similar retaliation against their activity, and the risk to the researchers should be considered at the project design stage. Research work involving engagement with the scambaiting community needs to account for the elevated risk profile of the participants, and design mechanisms to preserve anonymity. Sustained exposure to fraudsters can also carry a psychological cost to scambaiters~(\secref{sub:challenges}) and research engaging with the scambaiting community should account for this. Similarly, researchers engaging with scammers should take precautions such as limited engagement time and mandatory team rotations.

\subsubsection{Automation that Preserves Meaningful Participation}

The scambaiting workflow involves several stages of manual effort including sourcing targets, verifying numbers, and preparing and managing personas, which demand significant time given the volume of TSS scams~(\secref{sssec:time-workload-and-burnout}). Tools supporting scambaiters should automate repetitive preparatory tasks, such as sourcing phone numbers, while leaving higher-risk tasks such as verifying numbers to manual screening since misclassification could harm uninvolved parties. Given the scambaiting community's varying levels of technical expertise~(\secref{sssec:education-technical-back}), such tools should prioritize usability and accessibility alongside advanced functionality.

Additionally, while participants described their prosocial motivations~(\secref{sub:motiv-goal-bg}), they also valued the interaction and engagement aspects of scambaiting. Systems designed to support scambaiting should be careful that choices made in automation do not demotivate this highly-invested population by interfering with these community interactions.
This does not mean prioritizing the preservation of scambaiting as a hobby over other outcomes, but does suggest researchers should carefully consider the tradeoffs. Some types of automation may discourage engagement from experienced scambaiters who currently produce valuable, up-to-date adversarial intelligence, and this expertise may not easily be replaced.

\subsection{For Law Enforcement}
\label{sec:discussion.le}

Participants treated reporting to law enforcement as a pathway for investigation and prosecution, but also described difficulties identifying where useful intelligence should be submitted and frustration with delays in action~(\secref{sssec:delays_and_routing_gaps}). Law enforcement agencies should therefore provide clear and publicly accessible reporting channels specifying where reports should be submitted and what evidence is useful. Wherever possible, they should acknowledge the receipt of a report. Such acknowledgement could reduce frustration and preserve people's willingness to report scam activity~(\secref{sub:irresponsible}). In the US, the proposed ReportScams.gov Act (S. 4782) would move in this direction by establishing a centralized scam-reporting portal to route reports to appropriate authorities while providing reporters with confirmation, routing information, and follow-up mechanisms~\cite{reportscamsact2026}.

Investigators directly verifying reported scam operations should also account for scammers' screening tactics~(\secref{subsection:scammer_tactics}). Scambaiters' mechanisms for navigating call screening can help investigators maneuver past these checks when directly engaging suspected scam operations. More broadly, law enforcement can learn from scambaiters' investigative practices, including identifying scam operations and systematically collecting evidence for further investigation. With formal investigative powers, law enforcement agencies are well positioned to adapt such practices where legally and ethically appropriate, an opportunity also highlighted by Button et al.~\cite{buttonPolicingCrossborderFraud2025}. Rapid infrastructure rotation also makes timely action important. Reports containing short-lived indicators should be handled promptly and, where appropriate, coordinated with financial institutions and infrastructure and service providers to disrupt these operations while larger investigations continue.

Looking forward, beyond responding to reports, law enforcement could take a more proactive role in scam disruption by collaborating with researchers to operationalize tools and capabilities emerging from research on automated scam engagement and detection and mitigation systems for direct use in investigations. 

\subsection{For Policymakers}
\label{ssec:rec-policymakers}

TSS spans financial, technical, and legal domains, requiring responses from multiple stakeholders~(\secref{sssec:report-to-external-entities}). Policymakers should facilitate coordination among law enforcement, financial institutions, and infrastructure and service providers. At the federal level in the US, this need for coordination is also reflected in the proposed ReportScams.gov Act (S. 4782), which would establish an interagency Scams Steering Committee to coordinate federal anti-scam efforts~\cite{reportscamsact2026}. Where appropriate, policymakers and regulators can establish requirements to prevent and respond to the abuse of services used in scam operations~(\secref{sub:irresponsible}). Recent efforts in the UK to restrict caller ID spoofing~\cite{uk_phone_countermeasure}, along with US requirements for large voice service providers to implement caller ID authentication~\cite{us_phone_countermeasure}, illustrate how regulation can push infrastructure providers to adopt technical safeguards against the abuse described by participants.

Participants also viewed public education and victim support as important~(\secref{sub:awareness}). Policymakers should support accessible and up-to-date guidance describing common scam tactics, how potential victims should respond, and where incidents can be reported, as well as clear instructions for victims seeking help.

\subsection{For Financial Institutions}
Participants frequently prioritized reporting financial intelligence to banks because of their ability to act quickly, including freezing accounts or stopping transactions~(\secref{sssec:report-to-external-entities}). While some scambaiters had established direct contacts within banks, such access cannot be broadly available to everyone seeking to report. Therefore, financial institutions should provide clear reporting pathways for scam reports to ensure such information reaches appropriate fraud investigation departments. Financial institutions should also account for changes in how TSS scammers target and extract money from victims~(\secref{sub:evolving-scam-ecosystem}). Participants described scams shifting towards cryptocurrency. Current scam scripts and intelligence gathered through scambaiting can help financial institutions update their fraud detection systems and deliver effective customer warnings through their user interface.

\subsection{For Infrastructure and Service Providers}
In contrast to the proactive responses participants described from some financial institutions~(\secref{sssec:report-to-external-entities}), participants described infrastructure and service providers as slow or ineffective in responding to reported abuse~(\secref{sub:irresponsible}). Unlike financial institutions whose customers may be TSS victims, these providers' services may instead be used directly by scammers, providing less direct financial impetus to intervene. Nevertheless, they are positioned to directly disrupt reported resources, including phone numbers, domains, email accounts, and remote-access infrastructure. They should therefore maintain clear reporting channels and act promptly, particularly given the rapid rotation of scam infrastructure.

Beyond responding to reports, providers should also reduce opportunities for abuse proactively. Where incentives to address such abuse are weaker, policymakers can require providers to both prevent repeated misuse of their services and respond promptly to abuse reports. Furthermore, providers may also face reputational pressure when their infrastructure is repeatedly associated with scams, including through reports and discussions within scambaiting communities~(\secref{sub:evolving-scam-ecosystem}). ScreenConnect's changes to restrict customization features abused by scammers~(\secref{sub:evolving-scam-ecosystem}) provide one example of such platform-level mitigation. Broadly, more remote-access and online service providers should examine product features that scammers repeatedly exploit. Telecommunications providers should employ stricter verification for bulk phone-number acquisition and counter caller ID spoofing.

\subsection{Limitations} \label{subsection:limitations}
While our study provides valuable insights into the scambaiting community, it comes with several limitations that should be acknowledged. First, because we used semi-structured interviews, participants were not always asked the same follow-up questions, although all were asked the same key questions. Second, retrospective interviews are subject to recall issues, which can affect participants' memories of specific events. Real-time observational studies could reduce this limitation but would require researchers to embed themselves deeply in scambaiting practices. Third, participation was voluntary, so our sample may be affected by self-selection bias. We also rely on scambaiters' accounts of their own work and communities, without external perspectives other than those already captured in the literature. Participants may have presented their activities in a more socially desirable or ethically favorable manner. Although we used neutral questions and allowed participants to skip questions, self-reporting, over-reporting, under-reporting, and social desirability biases may remain. Our interviews also covered only legal scambaiting activities and participants did not discuss practices such as directly hacking scammer infrastructure~\cite{berneyNavigatingShadowsCyber2024}.

Finally, generalizability is not a typical expectation of qualitative research~\cite{leung2015validity}, and we do not claim that our findings represent the entire scambaiting community. Instead, we report the shared and differing views expressed by our participants. Although our sample included TSS scambaiters with varied backgrounds and experience levels, many came from overlapping communities, and our findings are closely tied to TSS scambaiting in particular. This focus provides targeted insight, but broader perspectives may require different recruitment and screening approaches.

\paragraph{\textbf{Future work.}} Future work engaging with scambaiters could focus on specific usability requirements for tools and reporting services that would measurably increase the disruptive effect and intelligence-gathering efficiency of scambaiting.
Additionally, future work could profit from a consolidation of scambaiting community resources. Scambaiting communities are constantly uploading archives of recorded calls, chat logs, and forum posts for anyone to see \cite{woodAnalysisScamBaiting2023, oak2025victims, noauthor_scammer_info_nodate, TechScammersUnitedScambaitForum, EaterWorlds419}. Future researchers may benefit from tracking this material to keep abreast of trends in scammer behavior.

\section{Conclusion}
\label{sec:conclusion}
Our analysis of 17 interviews reveals insights into how scammers deploy countermeasures against adversaries, how scambaiters prepare elaborate deceptive environments to evade suspicion, and how scambaiting activities still involve a significant manual burden. 
While a small number of prominent scambaiters perform these activities primarily for entertainment or public awareness, many others participate out of civic goodwill and a desire to protect potential victims. Many scambaiters actively collect, analyze, and report scam-related information, treating engagement with scammers as a means to gather intelligence and disrupt operations.
Collectively, these efforts constitute a distributed yet impactful form of community-driven defense that has, in practice, become one of the most significant responses to TSS. The collective experience and accumulated resources of this community thus represent a valuable resource for the research community.




\begin{acks}
We thank the anonymous shepherd and reviewers for their thoughtful feedback. We are especially grateful to Mo, Chairman of the Board of the Scammer.info forum, for helping us connect with members of the scambaiting community and arrange the interviews, and to the anonymous scambaiters who participated in our study and generously shared their experiences and insights. This work was supported in part by the National Science Foundation (NSF) under Grants CNS-2422035 and CNS-2544625 and by the Saudi Arabian Ministry of Defense.
\end{acks}

\bibliographystyle{ACM-Reference-Format}
\bibliography{references.bib}


\appendix 



\section{Open Science}
\label{sec:open_science}
To comply with the open science policy, we make the following artifacts publicly
available via our repository at \url{https://anonymous.4open.science/r/tss-interviews}

\textbf{Qualitative Codebook \& Transcripts:} The coding hierarchy, dictionary, and scripts used to calculate inter-rater reliability. We also include the text transcripts of the participants' responses. These artifacts can be accessed by downloading and opening the \texttt{coded\_segments.html} file in a browser.

\textbf{Artifacts Not Shared:}
As our research participants are scambaiters who operate in a highly adversarial space and may face potential retaliation from scammers, we do not share the raw video and audio recordings in accordance to our IRB protocols and consistent with the participant consent form.

\section{Ethical Considerations}
\label{sec:ethics} 


We took deliberate steps to ensure ethical conduct. As our study involved human subjects, we obtained IRB approval before beginning our study. At the start of each interview, participants were reminded of the study purpose, expectations, and withdrawal process and asked to confirm their consent to participate. As scambaiters may face retribution from scammers, we took measures to protect participant anonymity. Participants were instructed not to provide identifiable information in the recruitment form, to use pseudonyms on consent forms and during interviews, and to keep their cameras off. Audio recordings were stored in a secure, IRB-approved cloud storage accessible only to the researchers. We manually transcribed the recordings and removed or generalized any identifiable information. As stated in our IRB application and consent form, the audio recordings will be permanently deleted after publication of this paper. Participants were also informed that the anonymized transcripts and resulting findings would be retained and published. After publication, we will retain only participant pseudonyms and anonymized transcripts containing no identifying information.

Some members of the scambaiting community may be skeptical or reluctant to have their practices documented, as seen during participant recruitment~(\secref{ssec:screening-and-fraud-mitigation}). However, this reluctance cannot be generalized and does not appear to represent a community-wide consensus. For instance, during our recruitment, the founder and moderator of a prominent scambaiting community strongly emphasized the need to systematize scambaiter knowledge for research efforts, arguing the benefits outweigh the potential risks. The willingness of prominent scambaiters to participate in our study further suggests their openness to sharing their knowledge.

Apart from the human-subject protections, we recognize the dual-use risks associated with some of our recommendations. For instance, capabilities intended to support scambaiters and defensive efforts, such as persona realism tools and automated engagement, could also be repurposed by threat actors to improve their own practices. Researchers and practitioners should therefore consider these risks when deciding which capabilities to develop or release. Possible safeguards include restricting access to sensitive capabilities to vetted researchers and practitioners. For AI-assisted tools, researchers could also responsibly disclose high-risk findings and potential abuse cases to relevant AI providers so that they can strengthen safeguards against misuse.

\clearpage 
\section{Methodology Appendix}\label{sec:methodology_appendix}
\begin{table}[htbp]
    \caption{Breakdown of Coding Scope (Categories per Section)}
    \label{table:coding_scope}
    \centering
    \small
    \begin{tabular}{@{} p{0.3\columnwidth} p{0.65\columnwidth} @{}}
        \toprule
        \textbf{Section}    & \textbf{Permitted Categories}                                 \\
        \midrule
        1. Introduction     & motivation, sbr-bg, scammer-tactic                            \\
        2. Planning \& Prep & prep, goal, source, scam-verification, protection, legal      \\
        3. Tools \& Tech    & hardware, software, protection, vm-customization, cost        \\
        4. Engagement       & skill, payment, session-end, technique, intel, scammer-tactic \\
        5. Challenges       & challenge, session-end, scammer-tactic, trend, technique      \\
        6. Analysis         & success, logs, report, report-to, sb-system-advice            \\
        7. Best Practices   & protection, skill, sb-system-advice, automation               \\
        8. Conclusion       & sb-improvement, challenge                                     \\
        \bottomrule
    \end{tabular}
\end{table}

\begin{table}[htbp]
    \centering
    \caption{Distribution of Communication Channels Reported by Scambaiting Participants ($N=146$)}
    \label{table:channel_stats}
    \begin{tabular}{@{} p{0.6\columnwidth} p{0.10\columnwidth} r @{}}
        \hline
        \textbf{Communication Channel}                    & \textbf{Count} & \textbf{Percentage} \\
        \hline
        Phone Calls (Voice)                               & 91             & 62.3\%              \\
        Email                                             & 86             & 58.9\%              \\
        Text Messages (SMS)                               & 73             & 50.0\%              \\
        WhatsApp                                          & 71             & 48.6\%              \\
        Telegram                                          & 54             & 37.0\%              \\
        Instagram                                         & 48             & 32.9\%              \\
        Screen Sharing / Remote Desktop                   & 47             & 32.2\%              \\
        \textbf{Phone Calls \& Screen Sharing (Combined)} & \textbf{42}    & \textbf{28.8\%}     \\
        Dating Apps (Tinder, Bumble, etc.)                & 35             & 24.0\%              \\
        X (Twitter)                                       & 32             & 21.9\%              \\
        E-commerce (eBay, Facebook Marketplace)           & 30             & 20.5\%              \\
        Video Calls (Zoom, Skype, etc.)                   & 25             & 17.1\%              \\
        Selected ``All'' Options                          & 3              & 2.1\%               \\
        \hline
        \multicolumn{3}{l}{\textit{Note: Participants could select multiple channels.}}
    \end{tabular}
\end{table}


\begin{table}[htbp]
    \centering
    \caption{Participant Screening and Selection Process}
    \label{table:screening_process}
    \footnotesize
    \begin{tabular}{@{} p{0.25\columnwidth} p{0.6\columnwidth} r @{}}
        \hline
        \textbf{Strategy}         & \textbf{Description}                                   & \textbf{N}  \\
        \hline
        Strategy I:               & Total survey responses                                 & 146         \\
        Recruitment               &                                                        &             \\
        \hline
        Strategy II:              & Excluded: Missing ``Phone'' + ``Screen Sharing'' pair  & -104        \\
        Consistency               & Excluded: Selected ``All'' options (Manipulators)      & -3          \\
                                  & \textit{Participants passing strict checkbox criteria} & \textit{39} \\
        \hline
        Strategy III:             & Rescued: Valid text details                            & +20         \\
        Qualitative Override      & (e.g., ``ConnectWise'')                                &             \\
        \hline
        \textbf{Qualified Sample} & \textbf{Total Qualified Participants}                  & \textbf{59} \\
        \hline
        Dropouts                  & Excluded: Did not reply or unreachable                 & -37         \\
                                  & Excluded: Scheduled but did not attend                 & -4          \\
        \hline
        Interviewed Sample        & Participants interviewed from screening                & 18          \\
        \hline
        Post-Interview            & Excluded: Post-interview filtering                     & -2          \\
        \hline
        Valid Sample              & Included interviews from screening                     & 16          \\
        External Addition         & Special interview bypassing screening (P19)            & +1          \\
        \hline
        \textbf{Final Sample}     & \textbf{Interviews included in study}                  & \textbf{17} \\
        \hline
    \end{tabular}
\end{table}

Figure \ref{fig:forum_post_thread} illustrates the skepticism encountered during our recruitment process on public forums.
\begin{figure}[h]
    \centering
    \scampost{o}{Researcher (Original Poster)}{
        Hello all, \\
        We are a group of cybersecurity researchers from xxxxxxxx (redacted). We are conducting research on scambaiting and are looking for your help to participate in an interview. If you have experience with scam baiting are interested in helping us out, can you please fill out this screening form and distribute it to others: Microsoft Forms \\\\
        If you are selected for an interview, we will compensate you for your time by paying you \$30 USD. You can also stay anonymous during this process (by turning off your video) and giving us a pseudonym or fake name.
    }
    \scampost{u}{Member 1}{
        offering scambaiters a \$30 Gift card for an interview made me lol
    }
    \scampost{u}{Member 1}{
        but jokes aside, I looked at the form and you ask this:
        \begin{quote}
            \emph{``8. Do you interact with other scam baiters online? ... Please be as detailed as possible...''}
        \end{quote}
        What will happen to the data from the survey you create? Will you publish it? The reason why I ask cause this information is a double-edged sword as scammers read forums like these as well... \\
        Maybe you could give a little more background on the survey and why/how you are working on this topic. \\
        thanks
    }
    \scampost{o}{Researcher (Reply)}{
        The survey data will remain confidential and not be published in any identifiable form. Only general statistics, such as the number of participants, may be included in our findings. The survey is solely for selecting interview participants. Interviews will be recorded and transcribed in a privacy-preserving manner, and the full contents of the transcription will also be kept confidential. Any insights shared in our research will be fully anonymised to protect privacy and ensure no sensitive information is revealed. \\\\
        We appreciate your thoughtful question and share your concern about protecting the scam-baiting community. The purpose of our study is to better understand this community's needs, to help us in designing new openly-accessible tools or platforms that could magnify the effect of scam-baiting activity. If you have further questions or concerns, please reach out.
    }
    \scampost{u}{Member 2}{
        I dunno I'm Leary of this … but how do we really know??
    }
    \scampost{u}{Member 3}{
        What is the aim of this study. I know enough about this stuff to know that the reason they give you is often not the real reason (for obvious reasons)... i hope you will let us know whenever the results are published.
    }
    \scampost{u}{Member 4}{
        Sounds like a chod scammers, that's trying to mine information outta us…… He's trying to get ``Each And Everything'' from us!!!... \\
        Stinky Bhenchod!!!..…….
    }
    \Description{Snapshot of an anonymized forum thread showing users expressing skepticism and hostility towards a researcher's interview recruitment post.}
    \caption{Anonymized recruitment thread showing community skepticism.}
    \label{fig:forum_post_thread}
\end{figure}

\clearpage
\section{Interview Guide}
\label{appendix:interview-guide}

\noindent\textbf{Introduction}

\begin{enumerate}[
    label=\arabic*.,
    leftmargin=1.6em,
    itemsep=0pt,
    topsep=2pt,
    parsep=0pt,
    partopsep=0pt
]
     \item Can you tell me about your role as a scam-baiter and what motivated you to start scambaiting?
     \item How long have you been engaging in scambaiting activities, and what experience do you have in this field?
     \item How often do you engage in scambaiting?
     \item Do you focus on a particular type of scam? If so, what types do you target and why?
     \item What is your technical background? Do you have any experience with programming or cybersecurity?
\end{enumerate}

\noindent\textbf{Planning and Preparation}

\begin{enumerate}[
    label=\arabic*.,
    leftmargin=1.6em,
    itemsep=0pt,
    topsep=2pt,
    parsep=0pt,
    partopsep=0pt
]
    \item How do you plan and prepare before engaging with scammers?
    \item What factors influence your approach to a scambaiting session? \textit{(e.g., revenge, fun, research, community protection, etc.)}
    \item How do you typically gather the necessary information (phone numbers) about scammers before engaging with them?
    \begin{enumerate}[label=\alph*)]
        \item What resources or methods do you use to gather contact information?
        \item How do you verify that the contact information (phone numbers) is associated with scammers? 
        \item Do you gather information manually or use any automated processes to streamline this step?
    \end{enumerate}
    \item What communication channels do you primarily use to contact scammers? \textit{(e.g., via phone, email, social media, SMS, etc.)}
    \item What precautions do you take to avoid detection by scammers? \textit{(How do you avoid being detected as a scam-baiter?)}
    \item How do you consider legal and ethical implications in your scambaiting activities, and what steps do you take to ensure compliance with legal regulations in your jurisdiction?

\end{enumerate}

\noindent\textbf{Tools and Technologies}

\begin{enumerate}[
    label=\arabic*.,
    leftmargin=1.6em,
    itemsep=0pt,
    topsep=2pt,
    parsep=0pt,
    partopsep=0pt
]
    \item What hardware and software are essential to your scambaiting process?
    \textit{(To know how your scambaiting setup looks like.)}
    \item How do you manage the risk of scammers attempting to access your real personal information or systems? \textit{(How do you avoid being attacked, or harmed in anyway, by a scammer?)}
    \item What tools and technologies do you use to configure your system to appear like a genuine victim's system during an engagement?
    \item What are the costs associated with your scambaiting activities? \textit{(e.g., financial costs, time investment, resource allocation, etc.)}
    \item Do costs impact how you conduct scambaiting or the frequency of your engagements?
\end{enumerate}

\noindent\textbf{Engagement Process}

\begin{enumerate}[
    label=\arabic*.,
    leftmargin=1.6em,
    itemsep=0pt,
    topsep=2pt,
    parsep=0pt,
    partopsep=0pt
]
    \item In your experience, when do you think is the best time of the day to initiate a scambaiting session? And what days of the week are the best for scambaiting?
    \item Can you describe your strategy when engaging with a scammer? \textit{e.g., walk me through your typical process when engaging with a scammer over the phone.}
    \item How do you typically start a conversation with a scammer? What tactics do you use to keep them engaged?
    \textit{(starting with e.g., using certain keywords, tone, speech speed, speech volume, emotional state, etc.)}
    \textit{(tactics like e.g., pretending to be confused or elderly)}
    \item How do you respond when scammers ask for specific types of payment or information?
    \item When does a call usually end with a scammer, and how do you handle moments when they become suspicious?
    \item How do you gather information, or intelligence, about scammers during your engagements? \textit{(e.g., techniques, tools, or resources to gain more information about a scammer.)}
\end{enumerate}

\noindent\textbf{Challenges and Responses}
\begin{enumerate}[
    label=\arabic*.,
    leftmargin=1.6em,
    itemsep=0pt,
    topsep=2pt,
    parsep=0pt,
    partopsep=0pt
]
    \item What are the main challenges or pain points you face with scambaiting, and how do you overcome them?
    \item Do your calls ever fail? If so, why and how do they fail?
    \item Have you seen scammers changing their tactics over time? How do you adapt to these changes?
    \item In cases where scammers become aggressive or threatening, how do you de-escalate the situation?
    \item What tools or resources do you think could benefit the scambaiting process?
\end{enumerate}

\noindent\textbf{Analysis and Outcomes}
\begin{enumerate}[
    label=\arabic*.,
    leftmargin=1.6em,
    itemsep=0pt,
    topsep=2pt,
    parsep=0pt,
    partopsep=0pt
]
    \item How many calls do you make in a typical week (or month)? Does this number vary, and if so, what factors influence these variations? \textit{(e.g., special holidays might affect scammers' interactions.)} \item How do you define a successful call? \textit{(e.g., wasting time)}
    \item How long does a typical successful call last? Does this vary depending on the scammer's behavior?\item How frequently are you successful according to your definition of a successful scambaiting session? \item How do you collect and analyze data from your engagements to identify patterns or improve your strategies?
    \item What factors or outcomes would indicate that a scambaiting system is effective in combating scams? \textit{(What makes an effective scambaiting system?)}
    \textit{(factors like a scammer's behavior, the success rate, etc.)}
    \textit{(outcomes like preventing a scammer from achieving their goal.)}
    \item After a scambaiting session, how do you report the scammers, and to which authorities? \textit{(What do you do with your logs?)}    
\end{enumerate}

\noindent\textbf{Best Practice and Recommendations}
\begin{enumerate}[
    label=\arabic*.,
    leftmargin=1.6em,
    itemsep=0pt,
    topsep=2pt,
    parsep=0pt,
    partopsep=0pt
]
    \item Based on your experience, what best practices would you recommend to others interested in scambaiting?
    \item What skills or knowledge are essential for effective scambaiting?
    \item Do you have any advice for researchers or developers working on automated scambaiting systems?
    \item Have you used or are you aware of any automated tools or AI technologies in scambaiting?
    \item How receptive are you to integrating AI tools into your scambaiting activities?
\end{enumerate}

\noindent\textbf{Conclusion}
\begin{enumerate}[
    label=\arabic*.,
    leftmargin=1.6em,
    itemsep=0pt,
    topsep=2pt,
    parsep=0pt,
    partopsep=0pt
]
    \item How do you think scambaiting impacts the larger fight against technical support scams?
    \item Is there anything else you would like to share about your experiences as a scam-baiter that we have not covered?
    \item What Amazon market location would you prefer for your gift card? \textit{(US participants receive Tango Reward Link; Non-US participants receive region-specific Amazon gift card)}
\end{enumerate}

\clearpage 
\section{Codebook Appendix}\label{sec:codebook_appendix}
\begin{table}[htbp]
    \centering
    \small
    \caption{High-Level Codebook}
    \label{tab:codebook_appendix}
    \Description{Table showing categories and their respective codes used in the thematic analysis.}
    \begin{tabular}{@{} p{0.2\textwidth} p{0.75\textwidth} @{}}
        \hline
        \textbf{Category} & \textbf{Code}                                                                                                                                                                                                                                                                                                                                                                                        \\
        \hline
        automation        & ai-conversation-tools, ai-full-automation, non-ai-full-automation                                                                                                                                                                                                                                                                                                                                    \\
        \hline
        challenge         & maintaining-scammer-engagement, le-ineffectiveness, time-workload, public-awareness-of-scams, technical, persona-management, scammer-recognizes-baiter, scammer-op-barriers, reporting, misc                                                                                                                                                                                                         \\
        \hline
        cost              & telephony, hardware, software-services, sb-specialized-tools-suite, sb-time, none, misc                                                                                                                                                                                                                                                                                                              \\
        \hline
        hardware          & connectivity, computer-systems, telephony-infrastructure, audio-equipment, interface-controllers, props-and-aids, misc                                                                                                                                                                                                                                                                               \\
        \hline
        protection        & anonymity-and-identity-separation, telephony-Anonymity, session-control-mechanisms, malware-handling-and-analysis, device-and-environment-isolation, account-protections, network-protections, voice-and-audio-anonymity-safety-privacy                                                                                                                                                              \\
        \hline
        skill             & communication-and-social-persuasion, persona-performance-and-tactics, research-and-recon, vm-management-and-safety, it-and-scripting, telephony-systems-skills, offensive-reverse-connection-skills, call-management-principles                                                                                                                                                                      \\
        \hline
        scammer-tactic    & remote-access-and-system-control, social-engineering-and-manipulation, identity-masking, channel-migration, infrastructure-cycling-and-number-obfuscation, victim-acquisition-and-distribution, victim-verification-and-screening, demographic-targeting, scammer-technical-evasion-and-call-filtering, use-of-ai-and-emerging-technologies, localized-technical-demo-and-scareware, payment-tactics \\
        \hline
        motivation        & protect-others, anti-scam-sentiment, external-inspiration, personal-experience, recreational, profession-or-research                                                                                                                                                                                                                                                                                 \\
        \hline
        goal              & prevent-harm, personal-motivation, collect-and-report                                                                                                                                                                                                                                                                                                                                                \\
        \hline
        logs              & analysis-and-improvement, forensic-collection, peer-collab, note-logging, recording                                                                                                                                                                                                                                                                                                                  \\
        \hline
        vm-customization  & emulate-hardware, hide-vm-traces, create-usage-artifacts, pre-built-image, minimal-or-none                                                                                                                                                                                                                                                                                                           \\
        \hline
        report            & financial-account, contact-or-location, remote-access-credential, network-infrastructure-identifier                                                                                                                                                                                                                                                                                                  \\
        \hline
        trend             & media-feedback-loop, increased-scams, call-patterns-and-routing, script-evolution, payment-method-shift, infrastructure-or-tool-change, tech-evolution, misc                                                                                                                                                                                                                                         \\
        \hline
        report-to         & direct-contacts, banks, telecom-carriers, law-enforcement, organizations, sb-community, platforms-registries                                                                                                                                                                                                                                                                                         \\
        \hline
        technique         & de-escalate, indifference, confront, deflect-and-roleplay                                                                                                                                                                                                                                                                                                                                            \\
        \hline
        sb-improvement    & carrier-accountability, public-awareness-and-education                                                                                                                                                                                                                                                                                                                                               \\
        \hline
        handle-payment    & play-dumb, request-alternative-payment-method, hang-up, invent-excuse, provide-fake-payment                                                                                                                                                                                                                                                                                                          \\
        \hline
        intel             & prolonged-engagement, peer-sourced-information, open-source-lookup, insider-access, audio-and-context-cues, technical-capture                                                                                                                                                                                                                                                                        \\
        \hline
        legal             & compliance-and-no-intrusion, gray-area-and-risk, ethical-and-community-practice                                                                                                                                                                                                                                                                                                                      \\
        \hline
        prep              & persona-and-cheatsheet, voip-and-number-management, script-familiarization, vm-and-safety, honeypot-and-contact-collection, misc, audio-and-basic-checks                                                                                                                                                                                                                                             \\
        \hline
        session-end       & reveal-or-callout, goal-reached, payment-stalemate, suspicion-and-detection, tech-or-number-fail, human-error, misc                                                                                                                                                                                                                                                                                  \\
        \hline
        success           & intel-gathered, time-wasted, reported-or-enforced, scam-verification                                                                                                                                                                                                                                                                                                                                 \\
        \hline
        software          & deception-tools, telephony-tools, privacy-and-proxy, isolation-and-safety, intelligence-and-logging, audio-and-roleplay-support, research-and-osint, automation-and-scripting                                                                                                                                                                                                                        \\
        \hline
        scam-verification & active-calling, community-sourcing, carrier-and-osint                                                                                                                                                                                                                                                                                                                                                \\
        \hline
        source            & public-community, peer-network-and-private-sharing, passive-honeypot, web-scraping-and-osint, direct-or-personal-encounter, previliged-resources                                                                                                                                                                                                                                                     \\
        \hline
        sb-system-advice  & intel-and-disrupt, engagement-and-conversational-design, persona-and-voice, adaptation-and-learning                                                                                                                                                                                                                                                                                                  \\
        \hline
        sbr-bg            & formal-cs-edu, cybersec-pro, non-technical, self-taught-or-cert, it-pro-experience                                                                                                                                                                                                                                                                                                                   \\
        \hline
    \end{tabular}
\end{table}

\end{document}